\documentclass[aps,prx,twocolumn,longbibliography,superscriptaddress]{revtex4-2}
\usepackage{graphicx}
\usepackage{latexsym}
\usepackage{amssymb}
\usepackage{amsmath}
\usepackage{amsfonts}
\usepackage{upgreek}
\usepackage{bm}
\usepackage{multirow}
\usepackage{enumitem}
\usepackage{color}
\usepackage{hyperref}
\usepackage{physics}
\usepackage{bm}
\usepackage{youngtab}
\usepackage{ytableau}
\usepackage{float}
\usepackage{dsfont}

\usepackage{lipsum}
\usepackage{soul}

\hypersetup{
colorlinks = true,
linkcolor = [rgb]{0.70,0.13,0.13},
citecolor = [rgb]{0.13,0.55,0.13},
urlcolor = [rgb]{0.25, 0.41, 0.88}}

\newcommand{\U}{\mathrm{U}}

\newcommand{\ov}[1]{{\overline{#1}}}
\newcommand{\br}{{\bm{r}}}

\newcommand{\bR}{{\bm{R}}}

\newcommand{\bq}{{\bm{q}}}

\newcommand{\bG}{{\bm{G}}}

\newcommand{\bk}{{\bm{k}}}

\newcommand{\kivc}{\mathrm{KIVC}}

\newcommand{\eqnref}[1]{Eq.\,\eqref{#1}}
\newcommand{\figref}[1]{Fig.\,\ref{#1}}

\newcommand{\PRLsection}[1]{\emph{#1} -- }

\begin{document}

\title{Monte Carlo Studies of Twisted Bilayer Graphene: Strain and Thermal Fluctuations}

\author{Johannes S. Hofmann}
\affiliation{Department of Condensed Matter Physics, Weizmann Institute of Science, Rehovot 76100, Israel}
\affiliation{Max Planck Institute for the Physics of Complex Systems, N\"othnitzer Stra\ss e~38, 01187 Dresden, Germany}
\author{Patrick Ledwith}
\affiliation{Department of Physics, Massachusetts Institute of Technology, Cambridge, MA 02139, USA}
\author{Ashvin Vishwanath}
\affiliation{Department of Physics, Harvard University, Cambridge, MA 02138, USA}
\author{Jong Yeon Lee}
\affiliation{Physics Department, University of Illinois at Urbana-Champaign, Urbana, Illinois 61801, USA}
\affiliation{Korea Institute for Advanced Study, Seoul 02455, South Korea}
\author{Erez Berg}
\affiliation{Department of Condensed Matter Physics, Weizmann Institute of Science, Rehovot 76100, Israel}
\affiliation{Materials Department, University of California Santa Barbara, Santa Barbara 93106 USA}
\affiliation{Department of Electrical and Computer Engineering, University of California, Santa Barbara, CA 93106,
USA}

\date{\today}
\begin{abstract}
We study the phase diagram of twisted bilayer graphene at charge neutrality as a function of twist angle $\theta$, uniaxial heterostrain $\varepsilon$, and temperature $T$ using sign-problem-free quantum Monte Carlo simulations. At $T=0$ and zero strain, we find a continuous transition from a Dirac semimetal to a gapped Kramers inter-valley coherent (KIVC) phase as $\theta$ decreases toward the magic angle. With finite strain, the KIVC phase undergoes a further continuous transition at smaller $\theta$ into an anisotropic semimetal with gapless excitations near the center of the moir\'{e} Brillouin zone. In the KIVC regime, the entropy rises sharply with temperature and plateaus at $15\,\text{K} \lesssim T \lesssim 40\,\text{K}$ near the value expected from a Mott-like regime of localized electrons with nearly uncorrelated spin, valley, and orbital degrees of freedom, despite the topological obstruction preventing a localized tight-binding description of the active bands. 
The spectral function evolves continuously with $\theta$: at low $T$, a gap opens at the $K$ points and the minimal gap shifts to $\Gamma$ as $\theta$ decreases; at intermediate $T$, the spectral function smoothly interpolates between a Dirac semimetal spectrum with coherent $K$-point quasiparticles and a spectrum with gapless $\Gamma$-centered quasiparticles near the magic angle. The later can be a anisotropic or a Mott semimetal and we discuss how to distiguish them in experiment.
\end{abstract}
\maketitle

\PRLsection{Introduction}
Twisted bilayer graphene (TBG)~\cite{Li2010observation,MacDonald2011,Santos,Cao2018,Cao2018a,Balents2020,yang2023advance,bernevig2024twisted} offers a unique opportunity to study the interplay of electronic correlations and topology in a highly tunable setup~\cite{Efetov2019,Yazdani2019,CaltechSTM,Andrei2019,serlin2020intrinsic,sharpe2019emergent,Andrei2020,zondiner2020cascade,Wong2020cascade,xie2021fractional,yu2022correlated,grover2022chern,nuckolls2023quantum}. Tremendous theoretical effort has been devoted to understanding the electronic properties of TBG~\cite{Guinea2018,Po2018,Song18,Zou2018,Koshino2018,Yuan2018,Dodaro2018,PhilipPhillips2018,Tarnopolsky2019,Carr2019,Carr2019a,Po2019,Rademaker2019,Cea19,Seo2019,Sboychakov19,Thomson19,Cao2020,Xie2020,Wu2020,Bernevig2021,ledwith2021strong,Liu2021,Parker2021,Kwan2021,Shavit2021,Liu2021a,Wagner2022,Vafek2023,Breio2023,Sheffer2023,Zhu2024,Stubbs2025}. Particularly challenging is the intermediate coupling regime, where no theoretical expansion parameter is available. In this regime, numerically exact methods are invaluable, both for their predictive power and as a benchmark for other, approximate techniques~\cite{Haule2019,Kang2020,Potasz2021,Liao2021,Xie2021,Pahlevanzadeh2022,Xie2023,Wang2023,Zhang2023,Faulstich2023,Datta2023,Rai2024,Calderon2025,Xiao2025a}.

In this work, we employ numerically-exact determinant quantum Monte Carlo (DQMC) simulations to study the phase diagram of twisted bilayer graphene at the charge neutrality point. 
The single-particle part of the TBG continuum Hamiltonian is particle-hole symmetric to a good approximation. 
As shown in Refs.~\cite{Hofmann2022,Zhang2021momentum}, within this approximation the model describing TBG is free of the fermion sign problem at charge neutrality, allowing for large-scale DQMC simulations down to arbitrarily low temperatures. 
Moreover, the model remains sign-problem free in the presence of a number of physically relevant perturbations, such as heterostrain. 

\begin{figure}
    \centering
    \includegraphics[width=0.995\linewidth]{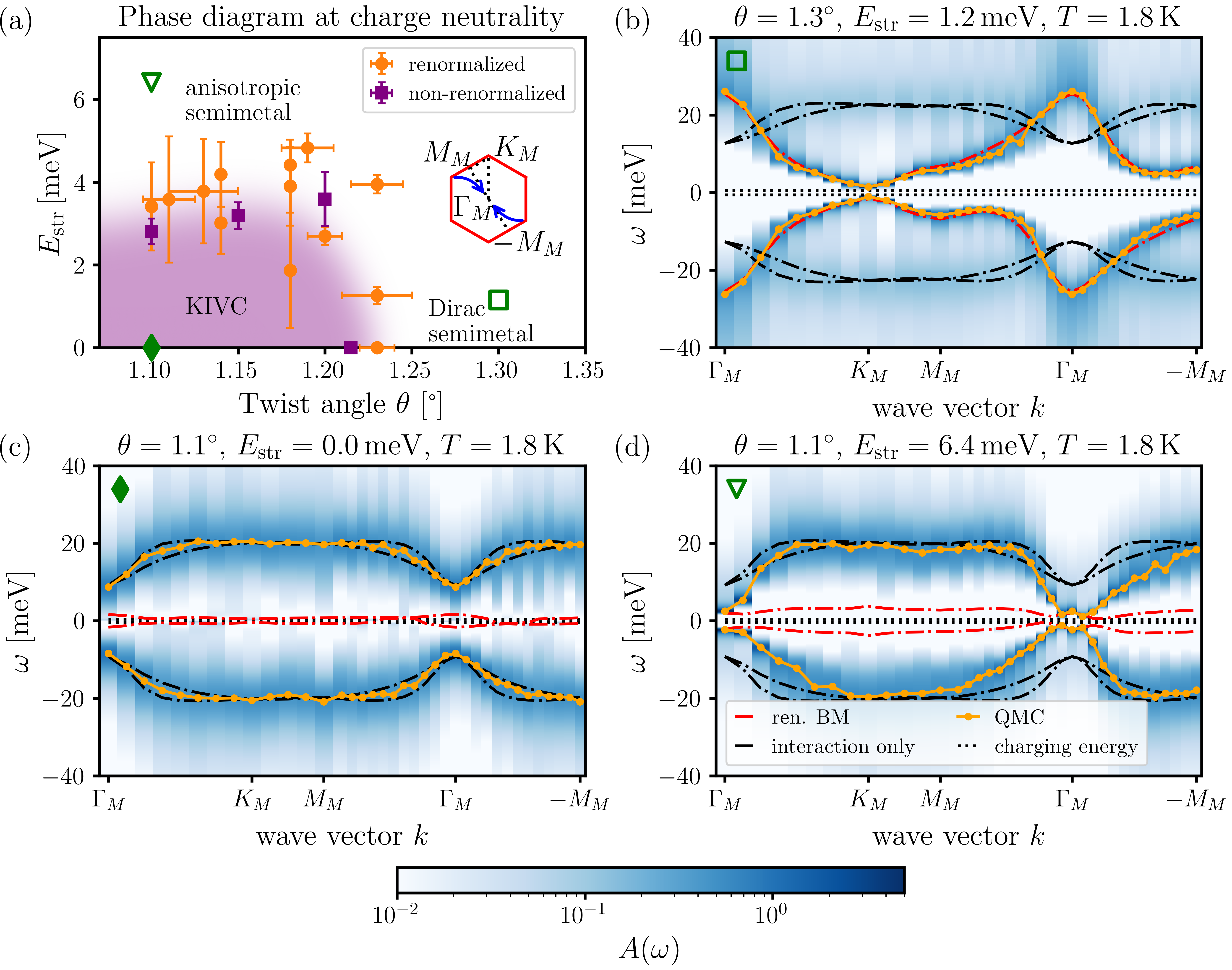}
    \caption{     
    {\bf (a) Phase diagram.} 
    A gapped KIVC phase (purple) borders an anisotropic semimetal (white), which is adiabatically connected to the $C_3$-rotation symmetric Dirac semimetal at large twist angle and vanishing strain. The green symbols indicate the parameters of panels (b-d). 
    The phase boundary was obtained using two different choices for the single-particle Hamiltonian, using the renormalized and non-renormalized schemes (see text). The non-renormalized data has been shifted to account for the magic-angle renormalization, $\Delta\theta=0.05^\circ$; the phase boundaries using the two schemes agree within error bars.
    The inset depicts the moir\'e Brillouin zone, the motion of the Dirac points (blue arrows) for $\theta=1.05^{\circ}$ upon increasing strain, and the high-symmetry path of the spectral plots (dotted line). 
    {\bf (b-d) Spectral function.} We plot $A(\bm{k},\omega)$ along the cut shown in the inset. Panel (b) corresponds to a point within the nearly-$C_3$ symmetric Dirac SM, panel (c) is within the KIVC phase, and (d) is within the anisotropic SM (d). The renormalized BM single-particle dispersion is shown in red. The black curves correspond to the single-particle excitation energies in an ``interaction only'' model where the single particle dispersion is artificially turned off ($\hat{H}=\hat{V}$). The orange curves correspond to the maximum of $A(\bk,\omega)$ obtained from DQMC. The black horizontal dotted line shows the charging energy due to the $\bq=0$ interaction in our $L=12$ system.
    }
    \label{fig:phase-diagram}
\end{figure}

Our results can be summarized as follows: (1) The $T\,{\rightarrow}\,0$ phase diagram as a function of twist angle, $\theta$, and uniaxial heterostrain, parametrized by the strain-induced energy scale $E_{\rm{str}}$,  shown in Fig.~\ref{fig:phase-diagram}(a), includes a continuous transition from the Dirac semimetal to a gapped KIVC phase at a critical value of $\theta$ at zero strain. At small non-zero $E_{\rm{str}}$, there is a further continuous transition at even lower $\theta$ back into a semimetal phase. In the small $\theta$, $E_{\rm{str}} \ne 0$  semimetal, the low-energy quasiparticles reside near the center of the moir\'{e} Brillouin zone ($\Gamma_M$), rather than near the corners of the Brillouin zone ($\pm K_M$). This phase has been identified in Hartree-Fock calculations~\cite{Liu2021,Parker2021,Kwan2021,Wagner2022} and dubbed the `nematic semimetal'; however, we stress that the phase we discuss here is found only for non-zero uniaxial heterostrain, and hence rotational symmetry is explicitly broken. We therefore refer to it here as the `anisotropic semimetal'.
(2) In the strain and angle regime where the ground state is a gapped KIVC state, there is a broad intermediate temperature range between $T\approx 15\text{K}$ and $T\approx 40\text{K}$ where the entropy is close to $k_B \ln{8 \choose 4}$ per unit cell. This is the value expected from a localized Hubbard-like picture, where charge fluctuations are quenched, and the entropy comes from the possible configurations of 4 electrons in 8 spin, valley, and orbital local levels. This behavior coexists with a gapless fermionic spectrum near $\Gamma_M$, as anticipated in Refs.~\cite{Haule2019,Datta2023,Rai2024,Ledwith2025,Hu2025}, and emerges even in the present model comprising exclusively of the flat bands~\cite{Hofmann2022, Ledwith2025}. 
(3) The fermionic spectral function displays a continuous evolution upon tuning $\theta$ at intermediate temperature from the Dirac semimetal, with coherent zero-energy quasiparticles at $\pm K_M$, to the vicinity of the magic angle, where spectrum is gapless near $\Gamma_M$. %\jyl{flat-band limit of KIVC state with band minimum at $\Gamma_M$ -- I think this is more general?}
We note that the gap closing at $\Gamma_M$ is related to the Berry-curvature concentration and not due to thermal broadening.

\PRLsection{Model and Method}
We study the continuum model~\cite{MacDonald2011} describing the two narrow moir\'e bands of TBG subject to uniaxial heterostrain with Coulomb repulsion. 
Focusing on charge neutrality and neglecting small terms in the Hamiltonian that break particle-hole symmetry~\footnote{Such particle-hole symmetry breaking terms contain those originating from the rotation of the Dirac Pauli matrices, proportional to $\theta$}, the model becomes free of the fermion sign problem~\cite{Hofmann2022,Zhang2021momentum}, enabling us to perform determinant QMC simulations at low temperatures and large system sizes. 

The model and the method are similar to those described in Ref.~\cite{Hofmann2022}, and will be briefly reviewed here (For more details see Appendix~\ref{app:details}). Since the narrow bands of TBG are topologically non-trivial, the calculation is performed in momentum space~\cite{Wang2021}, circumventing the need for a localized real-space basis. 
We introduce a new hybrid Monte Carlo variant (see Appendix~\ref{app:HMC}) which is more efficient than the Metropolis sampling used in Ref.~\cite{Hofmann2022}, and allows us to study larger systems, up to $L=12$.

Strain is incorporated into the TBG continuum model as described in Ref.~\cite{Balents2019general,Bi2019,Parker2021,Antebi2022}. In the presence of strain, the lattice is distorted, $\br \rightarrow (\mathds{1}+\mathcal{S}) \,\br$, and the strain tensor $\mathcal{S}$ for uniaxial strain is given by
\begin{equation}
    \mathcal{S} = \mqty( \varepsilon_{xx} & \varepsilon_{xy} \\ \varepsilon_{yx} & \varepsilon_{yy}) = {\cal R}_\phi \mqty( \varepsilon & 0 \\ 0 & -\nu \varepsilon) {{\cal R}_\phi}^T\, .
\end{equation}
Here, $\varepsilon$ is the strain magnitude, $\mathcal{R}_\phi$ is a rotation matrix, $\phi$ encodes the direction of the principle axis (we choose $\phi\,{=}\,0$ henceforth), and $\nu = 0.16$ is Poisson's ratio of graphene. With the change in the distance $d$ of carbon atoms the hopping amplitude $t(d)$ also varies, parametrized by $\gamma = -\frac{\partial \ln{t(d)}}{\partial \ln{d}}$. This effect of strain is represented by the pseudo gauge potential that couples with an opposite sign to the two valleys,
\begin{equation} \label{eq:gauge}
    \tilde{\bm{A}} = \tau_z \cdot \frac{ \gamma }{2a_0} ( \varepsilon_{xx} - \varepsilon_{yy}, -2\varepsilon_{xy})\,,
\end{equation}
where $\tau_z=\pm 1$ is the valley index. We use $\gamma=2$ throughout while varying the strain magnitude $\varepsilon$. 
We consider the case of heterostrain, $\varepsilon_1 = - \varepsilon_2 = \varepsilon/2$ (where $\varepsilon_{1,2}$ is the magnitude of the strain in the two layers), since the band structure of TBG near the magic angle is particularly sensitive to heterostrain~\cite{Koshino2018,Bi2019}. Importantly, within the continuum model, the pseudo gauge potential in \eqnref{eq:gauge} induced by strain does not break particle-hole symmetry, and hence the model is still sign-problem free~\cite{Hofmann2022}. 

The Coulomb interactions are projected to the two narrow bands of the BM model. For convenience, we use a dual-gated Coulomb interaction with a distance $d_0\,{=}\,20\rm{nm}$ between the system and the gates. A dielectric constant $\epsilon_r\,{=}\,10$ is used, representing screening from both the substrate and the remote bands. To avoid double-counting of the interactions, the Hartree-Fock subtraction scheme~\cite{Xie2020} has been proposed, subtracting from the single-particle Hamiltonian the Hartree-Fock potential evaluated for two decoupled graphene sheets. We use this renormalization scheme in some of the calculations and compare it to the unrenormalized scheme using the bare bands of the continuum model, as we discuss below.

\begin{figure*}
    \centering
    \includegraphics[width=0.995\linewidth]{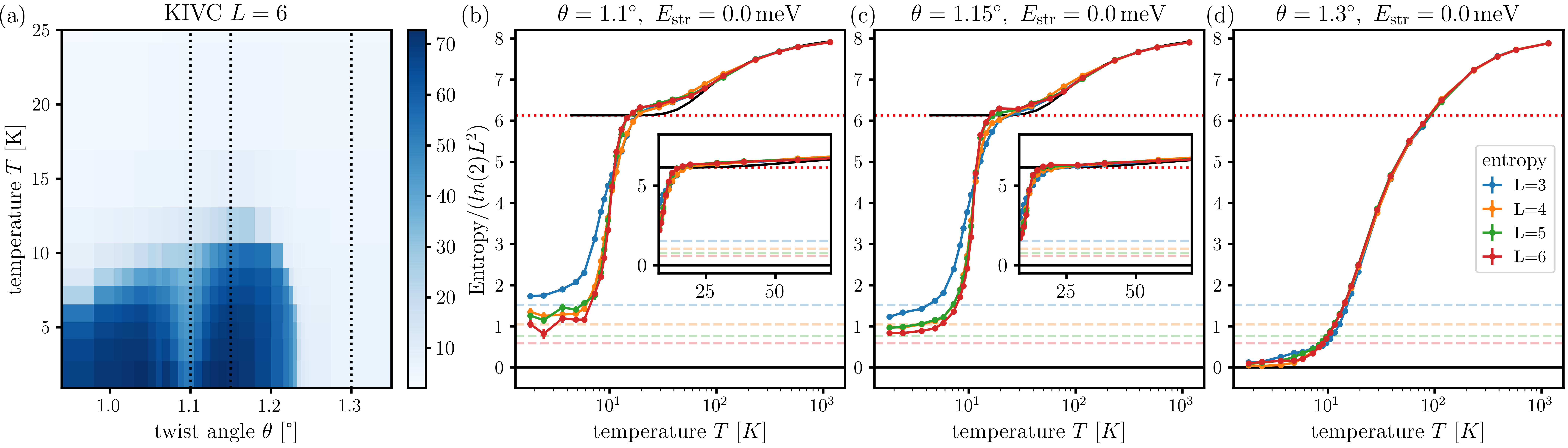}
    \caption{{\bf (a) Zero-momentum KIVC correlation function} for $\varepsilon=0$.
    At the magic angle, $\theta=1.1^{\circ}$, the single-particle bandwidth nearly vanishes, and the KIVC state is competing with valley polarized states due to an approximate $U(4)$ symmetry. 
    {\bf (b-d) Entropy per unit cell} as a function of temperature for three representative twist angles, (b) $\theta=1.1^{\circ}$, (c) $\theta=1.15^{\circ}$, and (d) $\theta=1.3^{\circ}$. The red dotted line at $S = \ln{8 \choose 4}$ represents the entropy of spin, valley, and orbital degrees of freedom for quenched electron on each site. The horizontal dashed lines represent the entropy expected for the degenerate ground states in the flat-band limit (see text). 
    In (b,c), the black curve represents the high temperature behavior obtained from a single-site SU(8) Hubbard model with $U=33 \rm{meV}$ and $U=35 \rm{meV}$ for $\theta=1.1^{\circ}$ and $\theta=1.15^{\circ}$. }
    \label{fig:entropy}
\end{figure*}

\PRLsection{Phase Diagram}
We begin by mapping out the low-temperature phase diagram of twisted bilayer graphene (TBG) at charge neutrality as a function of twist angle and heterostrain in Fig.\ref{fig:phase-diagram}(a). The amount of heterstrain is parametrized by $E_{\rm{str}}$, the energy splitting between the single-particle bands (renormalized by the Hartree-Fock subtraction) at the $\pm K_M$ points of the moir\'{e} Brillouin zone. At small strain, $E_{\rm{str}}\propto \varepsilon$. As we discuss in End Matter~\ref{EM:MeanderingDiracCone} (EM), $E_{\rm{str}}(\varepsilon)$ depends sensitively on details of the subtraction scheme; we comment further on the translation between $E_{\rm{str}}$ and $\varepsilon$ in the EM~\ref{app:end_matter}. We show the phase boundaries with (orange) and without (purple) band renormalization, which largely coincide.

The data used to construct the phase diagram was taken at $T=1.8\mathrm{K}$, which is sufficiently low to represent the $T\rightarrow 0$ behavior. 
The different phases are identified by measuring various fermion bilinear order parameters, as described in the EM~\ref{app:end_matter}, as well as the electron spectral function. The spectral function $A(\bm{k},\omega)$ along a cut in momentum space is shown at three representative points in the three different regions in the phase diagram in Fig.~\ref{fig:phase-diagram}(b-d).

We identify three distinct regions in the phase diagram. $(i)$ At large twist angles, we find a Dirac semimetal with two Dirac points near the $K_M$, $K_M'$ points of the moir\'{e} Brillouin zone. The spectral function in this region, extracted using the maximum entropy method~\cite{MaxEnt2004}, is shown in Fig.~\ref{fig:phase-diagram}b. At zero strain, the Dirac points are pinned to the $K_M$, $K_M'$ points by symmetry. $(ii)$ Upon decreasing $\theta$, the system undergoes a continuous phase transition into a Kramers inter-valley coherent (KIVC) phase~\cite{Bultinck2020,Hofmann2022}. For the definition of the KIVC order parameter, see~Eq.~\ref{eq:KIVC}. In this phase, the single-particle spectrum is gapped as can be seen in Fig.~\ref{fig:phase-diagram}c. 
$(iii)$ Increasing strain sufficiently close to the magic angle results in a further phase continuous transition in which the KIVC order parameter vanishes, and the system becomes an anisotropic semimetal, with a gap closing near the $\Gamma$ point of the moir\'{e} Brillouin zone, Fig.~\ref{fig:phase-diagram}d. 

We note that the anisotropic semimetal and the Dirac semimetal are not distinct phases: they can be adiabatically connected (see Fig.~\ref{fig:phase-diagram}a). Along this path, the Dirac points move continuously from the $K_M$, $K_M'$ points to the vicinity of the $\Gamma$ point. The trajectory of the Dirac points is shown schematically in the inset of Fig.~\ref{fig:phase-diagram}(a). We also note that heterstrain shifts the Dirac points away from zero energy, resulting in small electron and hole pockets in the anisotropic semimetal.

\begin{figure*}
    \centering
    \includegraphics[width=0.995\linewidth]{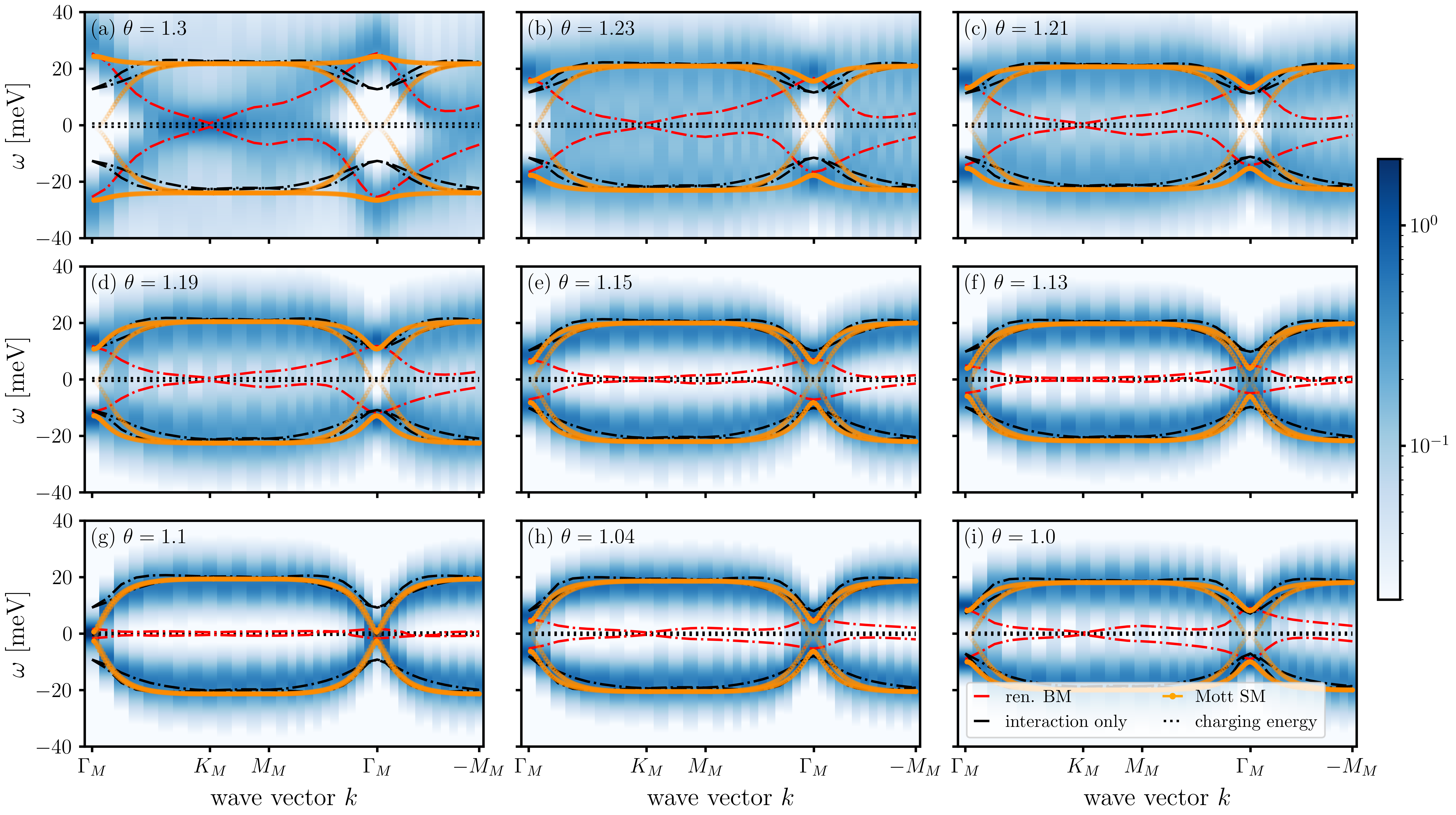}
    \caption{\textbf{Zero-strain single-particle dispersion} at various twist angles and $T=30\rm{K}$. The system size is $12\times12$. The renormalized dispersion, $H_0$, (red), the purely interacting spectrum, $H_0=0$, (black) and the charging energy, $V(\bq=0)$, (dotted line) are marked. The single-particle spectrum $A(\bk, \omega)$ is extracted via the stochastic Maximum-Entropy method. 
    The dispersion of the Mott semimetal is shown in orange with the opacity indicating the spectral weight.
    The dispersion smoothly evolves from a semimetal with Dirac cones at $K_M$ and $K_M'$ for large twist angle to a semimetal with gapless excitations at $\Gamma$ for the renormalized magic angle, $\theta=1.1^{\circ}$.}
    \label{fig:dispersion}
\end{figure*}

\PRLsection{Thermal Crossover and Entropy Plateau}
Next, we examine finite-temperature properties across the phase diagram. In the following calculations we used the renormalized band structure. Fig.~\ref{fig:entropy}(a) shows $\bm{q}=0$ KIVC correlation function versus temperature and twist angle at zero strain. Strong KIVC correlations onset below $\theta\approx 1.23^\circ$, with an onset temperature that reaches a maximum of $T_{\rm{KIVC}}\approx 10\rm{K}$ at $\theta \approx 1.15^\circ$. 
In our model, the finite-$T$ onset is a crossover rather than a true phase transition because the Hamiltonian has an exact $SU(2)\times SU(2)$ symmetry corresponding to independent spin rotations in each valley. Consequently, singlet and triplet KIVC states are exactly degenerate, and fluctuations on this enlarged order-parameter manifold preclude long-range order at any $T>0$ in two dimensions \footnote{In realistic samples, weak inter-valley Hund’s coupling breaks this symmetry and lifts the degeneracy, selecting either a singlet KIVC---which then exhibits a finite-temperature Berezinskii–Kosterlitz–Thouless transition---or a triplet KIVC, which remains disordered for all $T>0$.}.
Around $\theta\simeq1.10^\circ$, KIVC correlations drop while $T_{\mathrm{KIVC}}$ stays roughly constant. The drop reflects proximity to the flat-band ($W\to0$) regime with $U(4)$ symmetry~\cite{Bultinck2020}, where spin and valley ferromagnets are symmetry-related competitors to KIVC; in Appendix~\ref{app:AdditionalSpectrum} we shows other order parameters are enhanced in this angular window.

Fig. \ref{fig:entropy}(b-c) show the entropy per moir\'{e} unit cell versus temperature for twist angles $\theta=1.1^\circ$, $1.15^\circ$, and $1.3^\circ$. The first two angles are within the KIVC regime, whereas the third angle is in the Dirac semimetal phase. 
For $\theta=1.1^\circ,1.15^\circ$  and below $T_{\rm{KIVC}}\approx 10\rm{K}$ the entropy is small, and close to the value expected from the ground state degeneracy due to the approximate $U(4)$ symmetry for the corresponding system size (dashed horizontal lines). 
With increasing $T$, it rises rapidly and develops a broad plateau near $\ln \binom{8}{4}$ (horizontal dotted line), consistent with a regime in which charge fluctuations are quenched while the flavor (valley, spin, orbital) degrees of freedom are fully randomized---i.e., a ``flavor-incoherent'' state. 
This plateau persists up to $T\approx 40\rm{K}$, above which the entropy increases again, signaling the onset of charge fluctuations on each site.
For comparison, the solid black curve shows the entropy calculated from an atomic-limit Hubbard model with $\hat{H}_{\rm{HUB}} = \tfrac{U}{2} (\hat{N}-4)^2$, where $\hat{N}=0,\dots,8$ is the number of electrons on the site and $U$ is chosen to fit the entropy at high temperatures. Thus, within the KIVC phase and for $T\gtrsim10\rm{K}$, we find that the entropy is well-described by a Hubbard model in the limit of no hopping between the sites. This can be understood from the fact that, over much of the Brillouin zone, the narrow BM bands can be spanned by a basis of well-localized Wannier functions, and in this basis, the dominant interaction is the flavor-independent on-site Hubbard $U$ \cite{Wong2020cascade,Calderon2020,Song2022,Ledwith2025}.
On the other hand, in Fig.~\ref{fig:entropy}(d), the intermediate entropy plateau is absent in the Dirac semimetal phase. 
In Appendix~\ref{app:AdditionalEntropy} we show the entropy versus temperature in the presence of non-zero strain. As in the zero-strain case, a plateau at $S\approx \ln \binom{8}{4}$ appears, provided the strain is not so large as to destabilize the KIVC ground state. No such plateau is seen in the Dirac or anisotropic SM regions.

Finally, motivated by analytical progress on the spectral function in the flavor-incoherent regime~\cite{Ledwith2025,Hu2025,zhao2025mixed} and by recent measurements of the electronic dispersion~\cite{Xiao2025}, we compute the spectral function at $T=30\rm{K}$ and $\varepsilon=0$ for several twist angles (Fig.~\ref{fig:dispersion}). 
At $\theta=1.3^\circ$ the spectrum resembles that of the Dirac semimetal, with zero-frequency peaks at $K_M,K_M'$. 
Upon decreasing $\theta$ these peaks shift away to non-zero frequency, while the peak at $\Gamma_M$ shifts toward zero frequency, reaching a minimum near $\theta\approx1.1^\circ$ before drifting to higher frequency. Strikingly, the energy of the peak at $\Gamma_M$ closely tracks the single-particle Bistritzer–MacDonald dispersion (red curves), whereas away from $\Gamma_M$ the spectrum is nearly flat and gapped, with maxima near $\pm (15-20)\rm{meV}$\footnote{Strictly speaking, there is no true gap at $T>0$. By ``gapped'' we mean that the spectral function is peaked at a non-zero frequency.}. 

These features match recent analytical computations in the flat-band Hilbert space (\cite{Ledwith2025,ledwith2025a}, see also appendix), which leverage a small parameter $s^2$ that admits a systematic expansion. Here, $s$ characterizes the width of the Berry-curvature distribution of the topological flat bands. Related studies incorporating remote bands or additional orbitals\cite{ZhoiWannier,Po2019,Carr2019a,TBGII,Haule2019,Calderon2020,Song2022,Yu2023,Hu2023,Hu2023a,Herzogarbeitman2024}, treated either within DMFT~\cite{Datta2023,Rai2024,Calderon2020,calugaruObtainingSpectralFunction2025} or analytically~\cite{Hu2025,Lau2025,zhao2025ancillatheorytwistedbilayer,zhao2025mixed,vituri2026controlled,hu2026twisted,wei2026lifetime,nosov2026controlledexpansioncorrelatedelectrons}, have reported consistent results.

We note that while the spectrum may appear gapped at $\Gamma_M$, where the gap is set by the single particle dispersion, analytic calculations projected to the flat bands reveal a quadratic band touching with vanishing quasiparticle residue as $k\to 0$\cite{Ledwith2025,ledwith2025a}. These branches of the spectral function correspond to a trion excitation\cite{ledwith2025a} that becomes orthogonal to the electron as $k\to0$. Resolving this excitation would require computing the spectral function of the appropriate three-particle operator\cite{ledwith2025a}, which we leave for future work.

\PRLsection{Discussion}
Our findings establish a detailed phase diagram of TBG at charge neutrality using sign-free quantum Monte Carlo simulations. The identification of an anisotropic semimetal phase under strain, the characterization of the KIVC order and its melting with temperature, and the emergence of a high-entropy Mott-like regime provide a comprehensive understanding of the strongly correlated phases in this system.

Experiments in magic-angle TBG away from charge neutrality found a large entropy that appears at temperatures above a few degrees Kelvin, a significant part of which is of magnetic origin~\cite{Rozen2021,Saito2021}. This entropy has been attributed to strong spin and valley fluctuations associated with nearly localized states. Consistent with this view, we find that the entropy rises rapidly above the KIVC melting temperature of about $10\rm{K}$, approaching the value corresponding to completely random spin, valley, and orbital local moments. In the experiments, however, much of the entropy at $T\approx 10\rm{K}$ is quenched near charge neutrality. This suggests that strain stabilizes the semimetallic phase over much of the sample, since the anisotropic semimetallic phase has a significantly lower entropy at the same temperature range.

It is also instructive to compare our results with recent experiment~\cite{Xiao2025} measuring directly the momentum-resolved electronic spectrum of magic-angle TBG using the quantum twisting microscope (QTM)~\cite{Inbar2023}. At charge neutrality, the experiment found two gapped, nearly dispersionless bands at all momenta except near the $\Gamma_M$ point, where the gap closes. 
This dispersion qualitatively resembles both the spectrum of the anisotropic semimetal in the presence of strain (shown in Fig.~\ref{fig:phase-diagram}d at low $T$) and the Mott semimetal, realized in the flavor-incoherent regime (Fig.~\ref{fig:dispersion}, $\theta=1.1^\circ$). The temperature in the QTM experiment was $T=4\rm{K}$, well below our estimated onset of KIVC order in unstrained samples. 
Further QTM experiments at lower temperature and with varying degrees of strain (which is expected to vary across the sample) can establish the nature of the ground state at charge neutrality, i.e., cooling the Mott semimetal will lead to a gap while the anisotropic semimetal remains gapless down to much lower temperatures. 
The anisotropic semimetal exhibits (small) electron and hole Fermi surfaces; it is thus expected to be unstable at $T=0$~\cite{Grossman2023} towards an excitonic order, most likely some form of IVC. The Mott SM can also be differentiated from the anisotropic SM through a Zeeman field: while the bands of the anisotropic SM should simply spin split, increasing the density of states at low energy, the Mott SM instead responds by opening a gap $\propto U B_{\rm{Zeeman}}/T$\cite{Ledwith2025}, where $U$ is the interaction scale, at temperatures above the KIVC ordering scale.

\noindent{\bf Acknowledgments.} We are  grateful to Fakher Assaad, Lukas Janssen 
Eslam Khalaf, and Evyataar Tulipman for helpful discussions. This work was supported in part by NSF-BSF Grant DMR-2310312, by the European Research Council (ERC) under grant HQMAT (Grant Agreement No. 817799), Grant CRC 183 of the Deutsche Forschungsgemeinschaft (Project C02), a research grant from the Estate of Hermine Miller, the Sheba Foundation, and Dweck Philanthropies, Inc.,
the Simons Foundation Collaboration on New Frontiers in Superconductivity (Grant SFI-MPS-NFS-00006741-03), and the Deutsche Forschungsgemeinschaft DFG through the Würzburg-Dresden Cluster of Excellence on Complexity and Topology in Quantum Matter 'ct.qmat' (EXC2147, project ID 390858490). A.V. was supported by the Simons Collaboration on Ultra-Quantum Matter, which is a grant from the Simons Foundation (651440,
A.V.) 
We thank the hospitality of the Kavli Institute for Theoretical Physics (KITP) supported in part by grant NSF PHY-2309135, where part of this work was done.
The auxillary field QMC simulations were carried out with the ALF package~\cite{Assaad2022The2.0} available at \url{https://alf.physik.uni-wuerzburg.de}.

\bibliography{ref.bib}

\appendix

\newpage
\clearpage
\section{End Matter}
\subsection{Phase diagram vs. strain and KIVC correlation length}
\label{app:end_matter}

\begin{figure}[h!]
    \centering
    \includegraphics[width=0.995\linewidth]{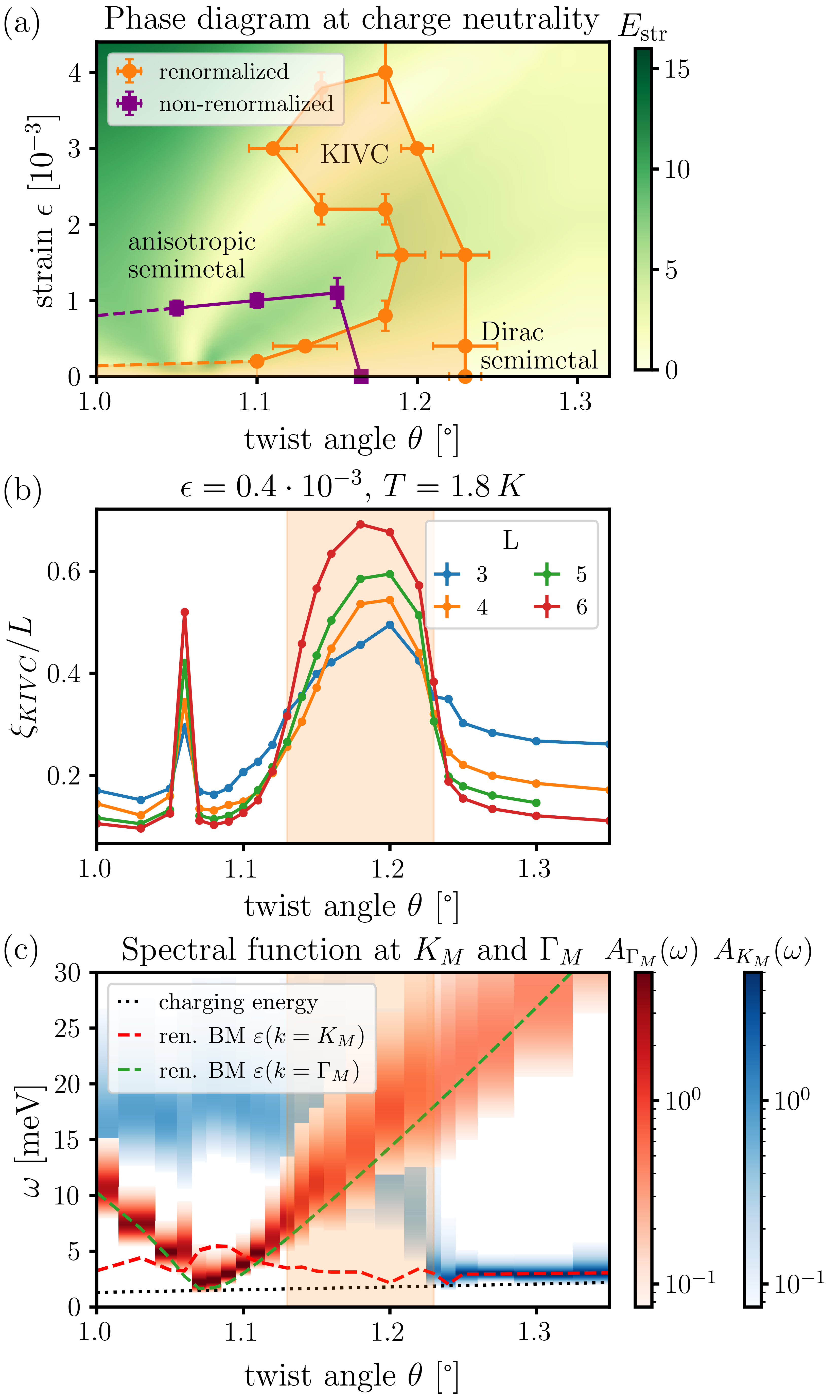}
    \caption{
    {\bf (a) Phase diagram.} 
    A gapped KIVC phase (orange) borders an anisotropic semimetal (transparent), which is adiabatically connected to the $C_3$-rotation symmetric Dirac semimetal at large twist angle and vanishing strain.  The dashed orange (purple) phase boundary for $\theta<1.1^\circ$ ($\theta<1.05^\circ$) is an extrapolation. The renormalized strain is shown as the density plot in the background, and translates the bare strain parameter $\varepsilon$ to the energy, $E_{\rm{str}}$, as shown in Fig.~\ref{fig:phase-diagram}(a).
    {\bf (b) KIVC correlation length.} $\xi_{\kivc}$ as function of $\theta$ for various system sizes at $\varepsilon=4 \cdot 10^{-4}$ and $T=1.8\rm{K}$. The crossing point locates the critical twist angles, $\theta=1.14\pm0.03^{\circ}$ and $\theta=1.23\pm0.02^{\circ}$, of the anisotropic SM to KIVC and KIVC to Dirac semimetal transition, respectively.
    {\bf (c) Spectral functions.} We show $A(\bm{k},\omega)$ at $K_M$ (blue) and $\Gamma_M$ (red) at $T=1.8\rm{K}$ as a function of twist angle $\theta$ and frequency $\omega$.
    }
    \label{fig:correl-length}
\end{figure}

In Fig.~\ref{fig:correl-length}(a), we present the same data as in Fig.~\ref{fig:phase-diagram}(a) of the main text, but as a function of the parameters $\theta$ and $\epsilon$ instead of the energy scales. The color map in green-yellow visualizes the strain energy scale, $E_{\mathrm{str}}(\theta,\epsilon)$, that we use to convert the two representations.

Phase boundaries are identified by the finite-size scaling. To illustrate this, in Fig.~\ref{fig:correl-length}(b), we show the KIVC correlation length, $\xi_{\mathrm{KIVC}}$ rescaled by the system size $L$ as a function of $\theta$ for different system sizes $L$ and for fixed strain, $\varepsilon = 0.4 \cdot 10^{-3}$.
It is extracted from the KIVC order parameter defined as~\cite{Bultinck2020}:
\begin{equation}
    \hat{O}^{x,y}_{\mathrm{KIVC}}(\bm{q})=\sum_{\bm{k},i,j} d_{i,\bm{k}}^\dagger d^{\vphantom{\dagger}}_{j,\bm{k}+\bm{q}}
    \langle u_{i,\bm{k}} \vert 
    \tau_{x,y}\sigma_x\mu_y \vert
    u_{j,\bm{k}+\bm{q}}\rangle,
    \label{eq:KIVC}
\end{equation}
Here, $d^{\dagger}_{i,\bm{k}}$ creates an electron in state $i = \{s,\tau,n\}$, corresponding to the spin, valley, and band indices, respectively, and momentum $\bm{k}$. $\vert u_{i,\bm{k}}\rangle$ is the periodic part of the Bloch state, and $\tau_\alpha$, $\sigma_\alpha$, $\mu_\alpha$ ($\alpha=x,y,z$) denote Pauli matrices acting in the valley, sublattice, and layer spaces, respectively. The definition of $\xi_{\rm{KIVC}}$ used in Fig.~\ref{fig:correl-length}(b) is given in Appendix~\ref{app:IVC}. 
The crossing points correspond to continuous phase transitions between the KIVC phase at $1.14^\circ \lesssim\theta \lesssim 1.23^\circ$ and the semimetallic phases outside of this range of $\theta$. The results are consistent with those of Ref.~\cite{Huang2025}, which found evidence for a continuous phase transition from the Dirac semimetal to a KIVC phase upon varying $\theta$ at $\epsilon=0$. Interestingly, at $\theta=1.06^\circ$ we also find enhanced KIVC correlations that increase with $L$~\footnote{This may be related to the fact that, near this twist angle and strain, the two Dirac points meet at $\Gamma_M$, which can enhance the tendency of the system towards KIVC order. More detailed calculations, including a finer $\theta$ grid and larger $L$, are needed to establish whether this is a true re-entrant KIVC phase.}. We use various scans (App.~\ref{app:IVC}) to map out the phase diagram and find a region of KIVC order, which coincides with a region of small $E_{\mathrm{str}}$.

Further insight into the nature of the phase transitions into and out of the KIVC phase is gained by examining the single-particle gap in the electron spectral function at the $K_M$ and $\Gamma_M$ points.  
Fig.~\ref{fig:correl-length}(c) shows the spectral function at the $K_M$ and $\Gamma_M$ points as a function of twist angle for the same strain as in panel (b). Clearly, at the $\theta\approx 1.23^\circ$ transition from the Dirac semimetal into the KIVC phase, a gap opens continuously at the $K_M$ point, whereas the transition at $\theta \approx 1.14^\circ$ from the KIVC to the anisotropic semimetal exhibits a gap closing at $\Gamma_M$. We also note that the gap at $\Gamma_M$ closely follows the single-particle dispersion of the renormalized BM model at the $\Gamma_M$ point (green dashed line).

\begin{figure*}
    \centering
    \includegraphics[width=0.94\linewidth]{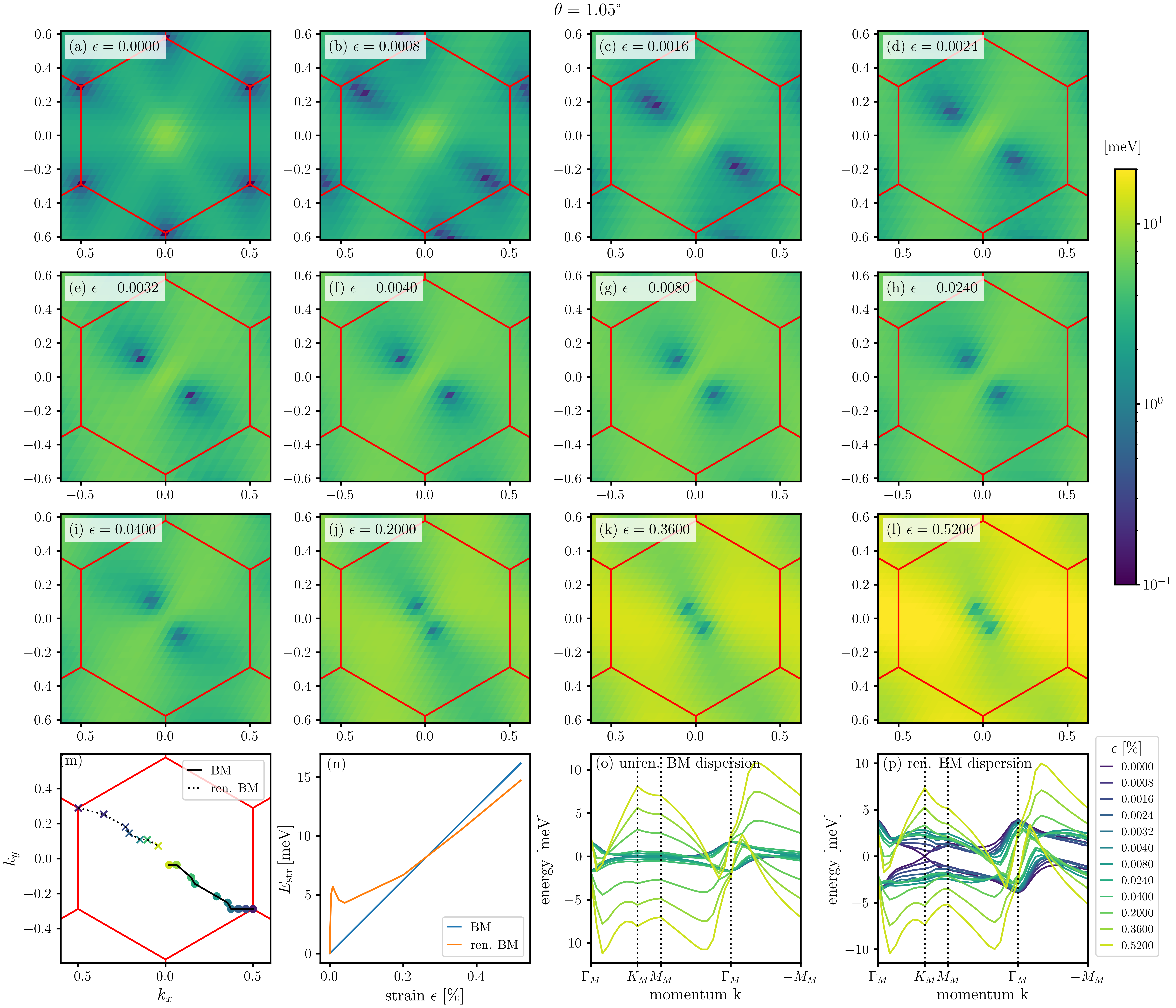}
    \caption{Meandering pair ($\mathcal{CT}$ symmetry) of the Dirac cones due to strain within the original moir\'e BZ: (a-l) gap $\Delta \epsilon(\bk)=\epsilon_2(\bk)-\epsilon_1(\bk)$ for various strain values; dark blue colors locate the Dirac cones within the moir\'e. (m) path of the Dirac points for the BM model (dots and solid line) and the renormalized dispersion (cross and dotted line). (n) band gap $\Delta \epsilon(K_M)$ as a function of strain. (o,p) BM- and renormalized dispersion along a high-symmetry path for a range of strain values.}
    \label{fig:meandering_DC}
\end{figure*}

\subsection{Renormalized and unrenormalized BM bands with heterostrain}
\label{EM:MeanderingDiracCone}

In this Appendix, we examine the single-particle bandstructure in the presence of strain, using either the bare (unrenormalized) BM continuum model or the renormalized BM bands where the 
the Hartree-Fock potential of the decoupled layers has been subtracted (see App.~\ref{app:details}). 
We find that the bands at a given amount of heterostrain $\varepsilon$ depend sensitively on the renormalization scheme. However, as shown in Fig.~\ref{fig:phase-diagram}, when the strain is expressed in terms of the energy scale $E_{\rm{str}}$ corresponding to the energy splitting between the bands at $K_M$, the phase diagrams obtained using the two schemes are almost identical.

The $C_3$ rotation symmetry pins the Dirac points to $\bk=\pm K_M$; for finite strain, which breaks this symmetry, the Dirac points generically meander within the BZ. In \figref{fig:meandering_DC}, we track their location as a function of $\varepsilon$ for both the bare and the renormalized BM model at fixed twist angle, $\theta=1.05^{\circ}$. Fig.~\ref{fig:meandering_DC}(a-l) depicts the band gap, $\Delta\epsilon^{\mathrm{BM,\,ren.}}(\bk)=\epsilon^{\mathrm{BM,\,ren.}}_2(\bk)-\epsilon^{\mathrm{BM,\,ren.}}_1(\bk)$, and the Dirac cones reside where the band gap vanishes (dark blue). In panel (m), we trace their motion from $\bk = K_M$ towards $\bk=\Gamma_M$ as a function of $\varepsilon$ for both the bare (solid dots) and the renormalized (crosses) BM model. Interestingly, the motion of the Dirac points in the renormalized model is considerably faster with $\varepsilon$ than in the bare model. This can also be seen in \figref{fig:meandering_DC}(n): even small values of $\varepsilon < 2\cdot 10^{-3}$ strongly increase the band gap $E_{\rm{str}} = \Delta \epsilon(K_M)$ compared to the bare BM dispersion.
Moreover, in the renomalized scheme, the dependence of $E_{\rm{str}}$ on $\varepsilon$ is non-monotonic, whereas in the unrenormalized scheme $E_{\rm{str}}(\varepsilon)$ is monotonic and nearly linear between $\varepsilon=0$ and $0.5\%$.

Panel (o) and (p) show the dispersion along a high-symmetry path for the original BM model and the renormalized bands, respectively. The unstrained BM model exhibits very flat bands at $\theta=1.05^{\circ}$. 
Hence, even weak strain can have a huge effect of the location of the Dirac points, especially near the magic angle where the zero-strain effective Dirac velocity vanishes. 

\newpage
\clearpage

\section{{Details of the Model}}
\label{app:details}
The single-particle Hamiltonian in the small twist-angle approximation reads 
\begin{align} \label{eq:tbg_ham}
    \hat{H}_\textrm{BM} &= \sum_{\bk}  c^\dagger_\bk H_\bk c_\bk + \sum_{\bk, \bk'} c^\dagger_{\bk'} T_{\bk',\bk}  c^{\,}_\bk   \nonumber \\
    H_\bk  &= \hbar v_F \big( k_x \sigma_x \tau_z + k_y \sigma_y  + \tilde{A}_x \sigma_x \mu_z + \tilde{A}_y \sigma_y \tau_z \mu_z \big)
    \nonumber  \\
    T_{\bk',\bk} & = \frac{1}{2} \sum_{n=0,1,2;\zeta=\pm 1} \delta_{\bk', \bk - \zeta \tau_z \bq_n}   \qty( \mu_x - i \zeta\mu_y ) T_n  \nonumber\\
    % T_{\bk',\bk} &= \frac{1}{2} \sum_n \delta_{\bk', \bk - \tau_z \bq_n}   \qty( \mu_x - i \mu_y ) T_n  + \textrm{h.c.} \nonumber\\
    T_n  &=  w_0   + w_1  e^{2\pi i n \sigma_z \tau_z /3} \sigma_x e^{-2\pi i n \sigma_z \tau_z /3}, %e^{-i \theta \sigma_z \mu_z \tau_z/2} dropped (small angle approx)
\end{align}
where $c^\dagger_\bk$ represents a row vector of electron creation operators with wavevector $\bk$ and with sublattice, valley, layer, and spin indices, denoted by $(\sigma,\tau,\mu, s)$, respectively. $v_F=8.7 \cdot 10^5 \rm{m/s}$ is the Dirac velocity [$\frac{2\hbar v_F}{3a_0}=2.7 \mathrm{eV}$, where $a_0 = 1.42 \textrm{ \AA}$ is the carbon-carbon distance in graphene]. $\theta$ is the twist angle, and $w_0=79\rm{meV}$, $w_1=105\rm{meV}$ are interlayer moir\'e hopping terms. The strain-induced pseudo-gauge potential $\tilde{\bm{A}}$ is given in Eq. (2) in the main text. The strain also distorts the moir\'e lattice vectors, $\bm{q}_n$ with $n=0,1,2$. Under the combination of rotation and heterostrain, 
\begin{equation}
    \bm{q}_n = (\mathcal{M}_1 - \mathcal{M}_2) \mathcal{R}_{2\pi n/3} \bm{K},    
\end{equation}
where $\mathcal{M}_\mu = (-1)^\mu (- \mathcal{S} + i\theta\sigma_y)/2$ is the combined rotation and strain tensor experienced by layer $\mu=1,2$ ($\mathcal{S}$ is the symmetric strain tensor given in Eq. 1 in the main text), $\bm{K} = (\tfrac{4}{3\sqrt{3}a_0},0)^T$ is the $K$ point of a reference undistorted and unrotated graphene lattice, and ${\cal R}_{\varphi} = e^{i \varphi \sigma_y}$ is a rotation matrix. 

In the second line of \eqref{eq:tbg_ham}, we have dropped the $\theta$ dependence in $H_\bk$ that originate from the rotation of the sublattice Pauli matrices by $e^{i\theta \sigma_z \tau_z \mu_z/4}$ that comes from the relative twist angle between the two layers. This amounts to dropping terms of order $\theta\approx 1.1^\circ \approx 0.02\,\rm{rad}$ at the magic angle. Under this approximation, the full action (including the interaction term, see below) is invariant under an anti-unitary particle-hole symmetry \cite{Hofmann2022}, 
\begin{align}
& \mathcal{C} c_\bk^\dagger \mathcal{C}^{-1} = c_\bk^T \tau_x \sigma_x \mu_y,\nonumber\\
& \mathcal{C}\, i\, \mathcal{C}^{-1} = -i.\label{eq:PHS}
\end{align}
This symmetry guarantees the absence of a sign problem at charge neutrality~\cite{Hofmann2022}.

To construct the Coulomb interaction projected to the narrow bands, we project the charge-density operator, with respect to the background at charge neutrality:
\begin{equation}\label{eq:denOps}
    \bar{\rho}_\bq = \sum_{\bk}  d^\dagger_{\bk + \bq} \Lambda(\bk,\bq) d^{\,}_{\bk} - \frac{1}{2} \sum_{\bk} \delta_{\bq, \bG}   \tr \Lambda(\bk,\bq)\,,
\end{equation}
We define $d^\dagger_{i,\bk}$ is a creation operator for a Bloch eigenstate of flavor $i=(s,\tau,n)$, i.e., spin $s$, valley $\tau$, and band $n$. $d^\dagger_{\bk}$ is a row vector containing the eight operators $d^\dagger_{i,\bk}$. The momentum $\bk$ is restricted to the $1^{\mathrm{st}}$ moir\'e BZ while the transfer momentum $\bq$ is unrestricted.
The form factor is defined as $\Lambda_{ij} (\bk,\bq) \equiv  \braket{u_{i,\bk+\bq}}{u_{j,\bk}}$, where $\vert u_{i,\bk} \rangle$ is the periodic part of the Bloch wavefunction.

To avoid the double-counting of the interactions, we subtract the Hartree-Fock mean-field contribution of a properly chosen reference state, whose corresponding single-particle density matrix is given by $P_0$. As discussed in the main text, we choose the reference state to be the Hartree-Fock ground state of two decoupled graphene layers~\cite{Xie2020}. The total Hamiltonian is then 
\begin{align}\label{eq:ham_QMC}
    \hat{H} &= \hat{H}_0  + \hat{V} \nonumber \\
    \hat{H}_0 &= \hat{H}_\textrm{BM} - \big[ \hat{V}\, \big]_{P_0}  \nonumber \\
    \hat{V} &= \frac{1}{2} \sum_\bq \bar{\rho}_\bq V_\bq  \bar{\rho}_{-\bq},
\end{align}
where $V_\bq = V_0 \tanh(\abs{\bq} d_0)/\abs{\bq}$ with $V_0 = \frac{e^2}{2 \epsilon_r \epsilon_0}\frac{1}{A}$ is the Fourier-transformed Coulomb potential with gate distance $d_0=20\mathrm{nm}$ and relative permittivity $\epsilon_r=10$. 

\section{Hybrid Monte Carlo method}\label{app:HMC}
The computational effort of the sign-problem free QMC methods, used in Refs.\cite{Hofmann2022,Zhang2021momentum}, scale with with the linear system size $L$ as ${(L^D)}^4$ (where $D=2$ is the spatial dimension), in contrast to the usual ${(L^D)}^3$ for determinant QMC simulations for lattice systems. This is due to the formulation in momentum space and the high rank, $\mathcal{O}(L^D)$, of the operators, $\rho_\bq$, that couple to the discrete auxiliary fields. To overcome this limitation, in the current work we use a hybrid Monte Carlo (HMC) method that generates global updates with a high acceptance rate \cite{Scalettar87,Kogut75,Beyl17,Lunts2023}. To this end, we use continuous auxiliary fields, $\phi$, and combine all interaction terms per time slice into a single exponential, akin to the Hubbard-Stratonovic decoupling of Coulomb interaction used in Ref.~\cite{Assaad2022The2.0},
\begin{align}
    \hat{B}_\tau &= \prod_{\bq,\zeta=\pm1} \left[ e^{-\Delta\tau V_\bq/8\,\zeta (\bar{\rho}_\bq+\zeta \bar{\rho}_{-\bq})^2} \right] e^{-\Delta\tau \hat{H}_0} \nonumber \\
    &\rightarrow e^{\sum_n  g_n\,\phi_{n,\tau} \hat{O}_n - \sum_n\phi_{n,\tau}^2/2}e^{-\Delta\tau \hat{H}_0}\,.
\end{align}
Here, $n=(\bq,\zeta)$ serves as an index for all interaction terms, $\hat{O}_n = \bar{\rho}_\bq+\zeta \bar{\rho}_{-\bq}$ at a fixed imaginary time $\tau$, and
$g_n = \sqrt{-\zeta\Delta\tau V_\bq/4}$ is the coupling constant.
Note that different $\hat{O}_n$'s typically do not commute, hence single field updates are inefficient. 
The particle-hole symmetry (Eq.~\eqref{eq:PHS}) also applies to the `effective' Hamiltonian \emph{after} the Hubbard-Stratonovich transformation. The density operator transform as $\mathcal{C}\bar{\rho}_\bq\mathcal{C}^{-1}=-\bar{\rho}_{-\bq}$ and since $V_\bq>0$, we have $\mathcal{C} (g_n,\hat{O}_n) \mathcal{C}^{-1}=-\zeta(g_n,\hat{O}_n)$ such that $g_n\hat{O}_n$ is invariant. Hence, the particle-hole symmetry guarantees the absence of the sign problem.

Within the HMC method, global updates are generated by following equal energy contours of an effective classical Hamiltonian,
\begin{equation}
    H_{\mathrm{HMC}} = \frac{1}{2} p^T M^{-1} p + V(\varphi)\,.
\end{equation}
Here, $\varphi=\{\phi_{n,\tau}\}$, is a vector containing all auxiliary fields, $p$ is a vector of the conjugate momenta to $\varphi$, $M^{-1}$ is inverse of the effective mass tensor with $M=\mathrm{diag}(1/g_n)$ and $V(\varphi)=-\log(\Tr{\prod_\tau \hat{B}_\tau})$ is the effective potential energy. The global updates are generated via hybrid molecular dynamics, by first drawing random conjugate momenta with probability $e^{-p^T M^{-1} p/2}$, and then integrate the equations of motion,
\begin{equation}
    \dot p = -\frac{\partial H_{\mathrm{HMC}}}{\partial\varphi}\quad\mathrm{and}\quad \dot\varphi=\frac{\partial H_{\mathrm{HMC}}}{\partial p}\,,
\end{equation}
using the leap-frog algorithm over an interval $t_{\mathrm{HMC}}$ where we typically use $N_{\mathrm{LF}}=30$ leap-frog steps with a discrete $\Delta t=0.05$ HMC time interval, i.e., $t_{\mathrm{HMC}}=30 \cdot 0.05$. Finally, we employ the usual Metropolis-Hastings scheme to accept or reject the proposed new configuration $\varphi(t_{\mathrm{HMC}})$. Note that $M^{-1}$, $\Delta t$, and $N_{\mathrm{LF}}$ are hyperparameters that determine the efficiency of the algorithm but do not affect any observables, which may be optimized even further to increase the acceptance rate using the largest possible leapfrog steps \cite{Lunts2023}.

Special care is required to evaluate the derivative with $\varphi$ due to the non-commuting operators combined to a single exponential:
\begin{align}
    \frac{\partial V(\varphi)}{\partial \phi_{n,\tau_i}} & = -\frac{1}{\Tr{\prod_\tau \hat{B}_\tau}} \frac{\partial}{\partial \phi_{n,\tau_i}} \Tr{\prod_\tau \hat{B}_\tau}\\
    & =- \frac{1}{\Tr{\prod_\tau \hat{B}_\tau}} \Tr{\prod_{\tau>\tau_i} \hat{B}_\tau \frac{\partial \hat{B}_{\tau_i}}{\partial \phi_{n,\tau_i}}   \prod_{\tau<\tau_i} \hat{B}_\tau }\,.
\end{align}
To proceed, we introduce the shorthand notation, $h(\varphi_{\tau_i})\equiv \sum_n g_n\phi_{n,\tau} \hat{O}_n - \sum_n\phi_{n,\tau}^2/2$, such that
\begin{align}
    \frac{\partial \hat{B}_{\tau_i}}{\partial \phi_{n,\tau_i}}   &= \frac{\partial e^{h(\varphi_{\tau_i})}}{\partial \phi_{n,\tau_i}}\\
    &= \int_0^1 dx \, e^{(1-x) h(\varphi_{\tau_i})}  \frac{\partial h(\varphi_{\tau_i})}{\partial \phi_{n,\tau_i}}  e^{x\,h(\varphi_{\tau_i})}\\
    &\approx \sum_i^{N_g} w_i\, e^{(1-x_i) h(\varphi_{\tau_i})}  \frac{\partial h(\varphi_{\tau_i})}{\partial \phi_{n,\tau_i}}  e^{x_i\,h(\varphi_{\tau_i})} \label{eq:gauss} \,,
\end{align}
where the second line is the continuous version of the product rule for derivatives and is valid also for non-commuting operators $\hat{O}_n$. The last line is discretizing the integral and we use the third-order Gaussian quadrature ($N_g=3$) with $w_i\in[5/18,8/18,5/18]$ and $x_i\in[\frac{5-\sqrt{15}}{10},\frac{1}{2},\frac{5+\sqrt{15}}{10}]$. Intuitively, one can think of $x_i$ as effective position in imaginary time \emph{within} a single Trotter time step. Combined, we have
\begin{widetext}
\begin{align}
    \frac{\partial V(\varphi)}{\partial \phi_{n,\tau_i}} & = - 
    \sum_j^{N_g} w_j
    \frac{
    \Tr{\prod_{\tau>\tau_i} \hat{B}_\tau 
    \, e^{(1-x_j) h(\varphi_{\tau_i})}  \frac{\partial h(\varphi_{\tau_i})}{\partial \phi_{n,\tau_i}}  e^{x_j\,h(\varphi_{\tau_i})}  
    \prod_{\tau<\tau_i} \hat{B}_\tau }
    }
    {\Tr{\prod_\tau \hat{B}_\tau}} \\
    &= \phi_{n,\tau_i} - 
    g_n
    \sum_j^{N_g} w_j
     \frac{
    \Tr{\prod_{\tau>\tau_i} \hat{B}_\tau 
    \, e^{(1-x_j) h(\varphi_{\tau_i})}  \hat{O}_n  e^{x_j\,h(\varphi_{\tau_i})}  
    \prod_{\tau<\tau_i} \hat{B}_\tau }
    }
    {\Tr{\prod_\tau \hat{B}_\tau}} \\
    &= \phi_{n,\tau_i} - 
    g_n
    \sum_j^{N_g} w_j
     \langle\langle \hat{O}_n(\tau_i+x_j) \rangle\rangle_\varphi
    \,,
\end{align}
\end{widetext}
where $\langle\langle \hat{O}_n(\tau_i+x_j) \rangle\rangle_\varphi$ refers to the expectation value of the operator $\hat{O}_n$ that couples to $\phi_{n,\tau_i}$ for a \emph{given} auxiliary field configuration $\varphi$ \emph{within} the Trotter interval $\tau_i$.

\begin{figure*}
    \centering
    \includegraphics[width=0.995\linewidth]{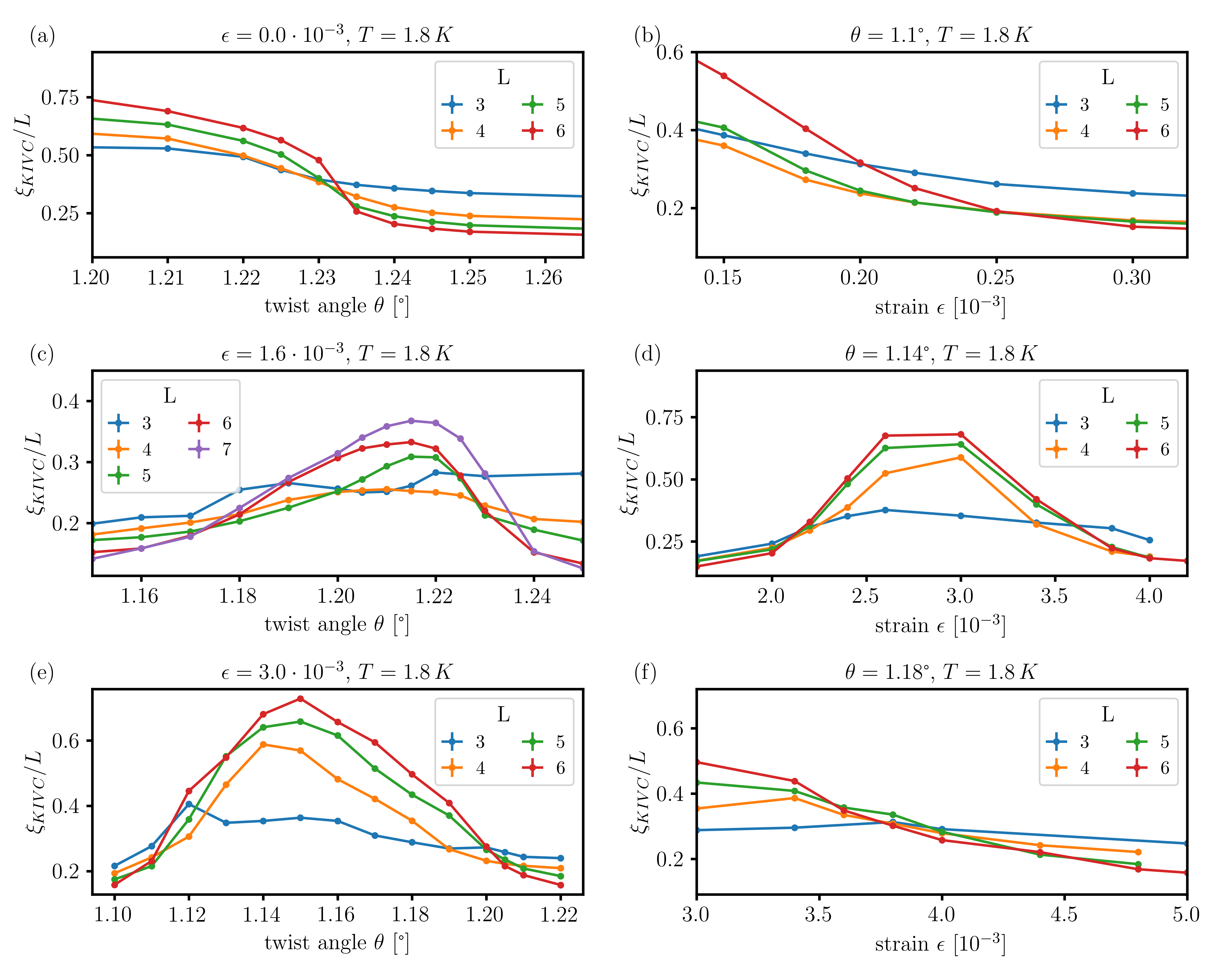}
    \caption{Additional data for the KIVC correlation length $\xi_{\mathrm{KIVC}}$ at $T=1.8\rm{K}$ used to determine the phase diagram of Fig. 1(a). The crossing points are used to extract critical twist angle (left column) or strain (right column) values.}
    \label{fig:correllength}
\end{figure*}

After each leap-frog iteration, the proposed update $\varphi(t=0)\equiv\varphi_i \rightarrow \varphi(t_\mathrm{HMC})\equiv\varphi_f$ is accepted while maintaining detailed balanced. The acceptance probability is $p_{acc}=\min(1,r)$ with
\begin{align}
    r &= \frac{P(\varphi_f \rightarrow \varphi_i)}{P(\varphi_i \rightarrow \varphi_f)} \frac{\Tr{\prod_\tau \hat{B}_\tau(\varphi_f)}}{\Tr{\prod_\tau \hat{B}_\tau(\varphi_i)}}\,,
\end{align}
where $P(\varphi_f \rightarrow \varphi_i)=e^{-p_f^TM^{-1}p_f/2}$ is the proposal probability of the update. Note that this requires the integrator of the equations of motion to conserve time reversal symmetry. That is, if the initial configuration $(\varphi_i,p_i)$ integrates to the finial configuration $(\varphi_f,p_f)$, then running the HMC algorithm from $(\varphi_f,-p_f)$ has to generate $(\varphi_i,-p_i)$. This is a property of the leap-frog algorithm and this time-reversal symmetry is maintained by the Gaussian quadrature of the derivative. 
We would like to stress that neither the finite step size $\Delta t$ nor the discretization of \eqnref{eq:gauss} introduce any systematic error as they are taken into account by `detailed balance'. Both approximations lead to violations of the energy conservation and this is accounted for by the acceptance ratio, $r=e^{-\Delta E_\mathrm{HMC}}$ with $\Delta E_\mathrm{HMC}=H_{\mathrm{HMC}}(\varphi_f,p_f)-H_{\mathrm{HMC}}(\varphi_i,p_i)$.

Finally, the numerically expensive tasks are calculating $e^{h(\varphi_{\tau_i})}$, which involves diagonalizing $h(\varphi_{\tau_i})$ and the matrix multiplication to evolve the equal-time greens function. These operations scale with system size $(L^D)^3$ such that the overall numerical expense is $\mathcal{O}(N_{\mathrm{Trot}}N_g(L^D)^3)$ similar the the usual auxiliary field QMC algorithm.

\section{KIVC correlation length}
\label{app:IVC}

We determine the phase boundaries of the KIVC order by analysing the associated correlation length. The projected operator $\hat{O}^{x,y}_{\mathrm{KIVC}}(\bm{q})$ defined in Eq. (3) of the main text. 
The associated correlation function, $S(\bq)$ is defined as
\begin{align}
    \label{eq:order_parameter} 
    C_\kivc(\bq) &= \frac{1}{L^2}\sum_{a=x,y}\langle \hat{O}^{a}_{\mathrm{KIVC}}(\bq) \hat{O}^{a}_{\mathrm{KIVC}}(-\bq) \rangle
\end{align}

\begin{figure*}
    \centering
    \includegraphics[width=0.995\linewidth]{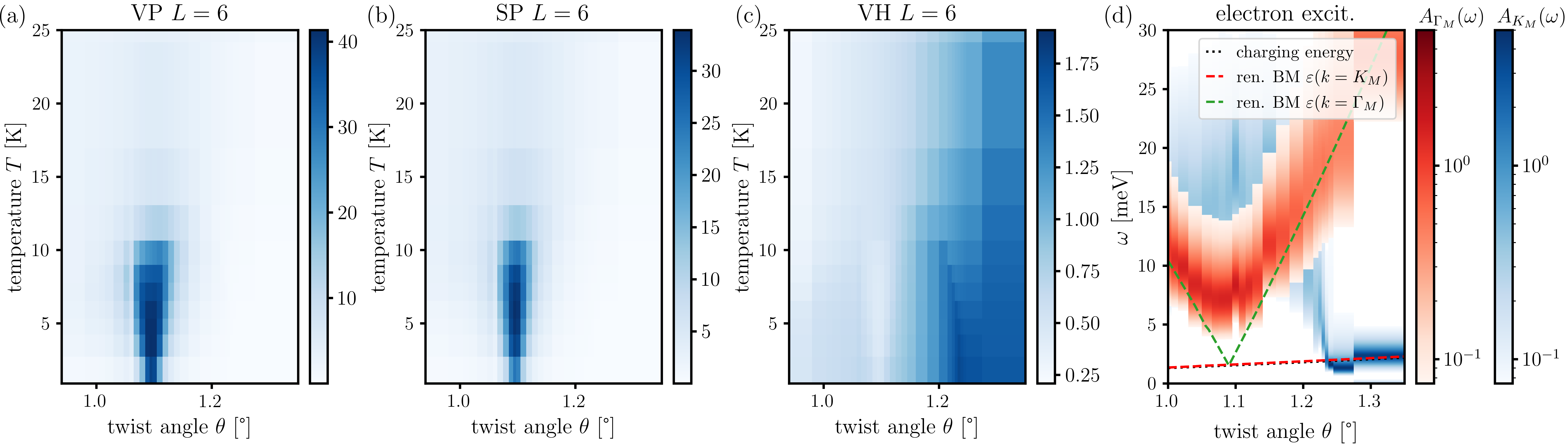}
    \caption{Correlation functions at $\bq=0$ in the absence of strain ($\varepsilon=0$) for valley  (a), spin  (b), and valley Hall (c) order parameters. The valley and spin polarized states compete with the KIVC order around $\theta=1.1^{\circ}$. The valley Hall state is disfavored throughout the phase diagram. (d) Evolution of the single-particle spectrum at $K_m$ (blue) and $\Gamma_M$ (red) with twist angle for $\varepsilon=0$ and $L=6$. We find a continuous gap opening at $\bk=K_M$ at $\theta_c=1.22^{\circ}\pm0.01^{\circ}$. The dispersion at $\bk=\Gamma_M$ mostly follows the expectation from the renormalized BM model (dashed green line), but remains gapped around $\theta=1.1^{\circ}$ (inside the KIVC state). The dashed red line denotes the finite-size charging energy.}
    \label{fig:U4-symmetry}
\end{figure*}

The correlation length, $\xi_{\kivc}$, is related to the curvature of the Bragg peak at $\bq=0$ for homogeneously ordered phases. For a triangular lattice, it can be calculated as \cite{Toldin15}
\begin{equation}
\xi^2_{\kivc}/a^2_M=\frac{3}{16\sin^2(\pi / L)} \left(\frac{C_\kivc(\bq=0)}{\overline{C_\kivc}(\bq=\bq_{nn})}-1\right),\label{eq:CorrLength}
\end{equation}
where $\bq_{nn}$ are the six nearest neighbor momenta to $\bq=0$ and $\overline{C_\kivc}(\bq=\bq_{nn})$ is averaged over those. 

In \figref{fig:correllength}, we show additional data for the ratio $\xi_{\kivc}/L$ for various scans with twist angle and strain. The crossing points of different system sizes locates the critical values between regions with KIVC order. This data was used to determine the phase boundary shown in Fig.~1(a) of the main text.

\section{Additional correlations and spectral data}
\label{app:AdditionalSpectrum}

For strain-free TBG, we show the correlation functions for homogeneous ($\bq=0$) valley polarization, spin polarization and valley-Hall order (corresponding to the order parameters $\tau_z$, $s_z$, and $\sigma_z$, respectively) in \figref{fig:U4-symmetry}(a-c) for various twist angle and temperatures. As mentioned in the main text, there is a strong enhancement of valley and spin correlations for a narrow twist-angle range around $\theta=1.1^{\circ}$, where the kinetic energy is maximally quenched. We stress the correlation function peak around $T\sim 5\mathrm{K}$ and weaken again at the lowest temperatures. TBG exhibits a $U(4)$ in the flat-band limit where the KIVC and the spin/valley polarized states are degenerate, while a finite kinetic energy favors the KIVC order, in line with our data. The valley Hall state is degenerate with KIVC in the flat band chiral limit~\cite{Bultinck2020} (where the intra-sublattice hopping term $w0$ is set to zero), 
but is disfavored in our calculation since we use $w_0/w_1=0.75$.

The evolution of the single particle spectrum at $\varepsilon=0$ with varying twist angle is shown in \figref{fig:U4-symmetry}(d) for $\bk=K_M$ (blue) and $\bk=\Gamma_M$ for $L=6$. At large $\theta$, the excitation energy at the Dirac point follows the dispersion of the renormalized BM model shifted by the charging energy due to the Coulomb interaction with vanishing transfer momentum $V_{\bq=0}$. This charging energy is a finite size effect and vanishes in the thermodynamic limit. 
Upon reducing the twist angle below $\theta=1.22^{\circ}$, a gap opens at the Dirac point $\bk=K_M$ continuously. Around $\theta=1.1^{\circ}$, the excitation energy at $\bk=\Gamma_M$ deviates from the renormalized BM value and reaches its minimum, but remains finite, i.e., above the charging energy. Hence, the single-particle spectrum is gapped within the KIVC phase. This is in contrast to the spectrum of the anisotropic semi-metal large enough strain (Fig.~2b of the main text) that exhibits gapless excitations in the center of the BZ.

\begin{figure*}
    \centering
    \includegraphics[width=0.995\linewidth]{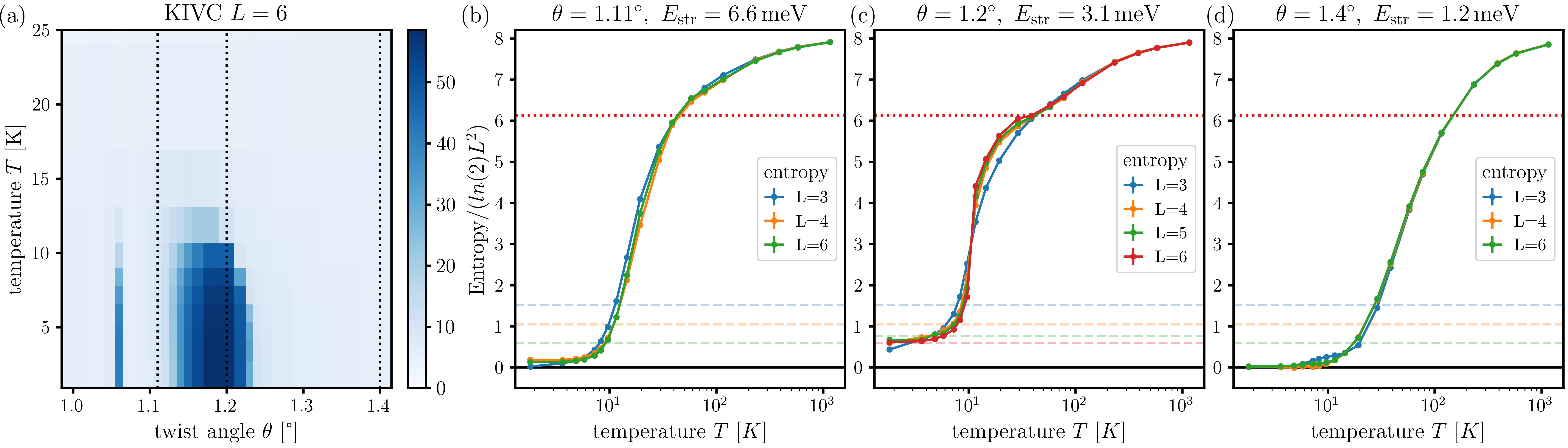}
    \caption{(a) Finite temperature phase diagram at non-zero strain, $\varepsilon=4 \cdot 10^{-4}$. The color indicates the $\bq=0$ KIVC correlation function, $S_{\kivc}(\bq=0)$, showing a region with significant KIVC correlations around $\theta=1.2^{\circ}$ and another narrow region at $\theta=1.06^{\circ}$ where the meandering Dirac cones meet at the $\Gamma_M$-point. (b-d) Entropy per unit cell with the same reference values as Fig.~3. For $\theta=1.2^{\circ}$ with a KIVC ground state, panel (c), a small plateau around $T\approx30\rm{K}$ is visible, consistent with quenched charge fluctuations.}
    \label{fig:entropy_ws}
\end{figure*}

\section{Entropy calculation and additional data}
\label{app:AdditionalEntropy}

To extract the entropy $S$ in DQMC simulation, 
we use the relation $S=-\left(\frac{\partial \Omega}{\partial T}\right)_{A,\mu}$, where $\Omega=k_\mathrm{B}T\ln{Z}$ is the grand canonical potential and the derivative is taken at fixed chemical potential, $\mu$, and area, $A$. Hence, we have $S=k_\mathrm{B}\ln{Z} + \langle H \rangle/T$, however, we cannot directly estimate $\ln{Z}$ in DQMC simulations\footnote{While $Z$ determines the relative weight of configurations, its normalization is unknown and not required for detailed balance.}.

Instead, we take another derivative with respect to temperature, and integrate along a path to determine the change in entropy,
\begin{align}
    \frac{\partial S}{\partial T} & = T^{-1} \frac{\partial}{\partial T} \langle H \rangle \\
    S(T_1) & = S(T_0) + \int_{T_0}^{T_1} dT\, T^{-1} \frac{\partial}{\partial T} \langle H \rangle\,.
\end{align}
Note that this is the well-know relation between the entropy change and the specific heat. As the latter is numerically challenging to estimate precisely in DQMC, we integrate the last term in the second line by parts and charge variable from the temperature to the inverse temperature $\beta$,
\begin{align}
    S(\beta_1) = & S(\beta_0) + k_\mathrm{B} \beta_1\left(\langle H \rangle_{\beta_1} - \langle H \rangle_{\beta_0}\right) \nonumber \\
    & - k_\mathrm{B} \int_{\beta_0}^{\beta_1} d\beta\, \left( \langle H \rangle_\beta -\langle H \rangle_{\beta_0}\right)\,.
\end{align}
A convenient choice of the reference state for the integration is the infinite temperature state with $\beta_0=0$ and $S(\beta_0)=8 k_\mathrm{B}L^2\ln{2}$.

Next, we determine the ground-state degeneracy of the KIVC ordered phase with perfectly flat bands, making use of the enhanced $U(4)$ symmetry. That is, we identify one reference state and then construct the ground-state manifold via $U(4)$ symmetry operations. As the reference state, we pick a fully valley polarized state which preserves the global $SU(2) \subset U(4)$ spin rotation symmetry. This state is a product state which is symmetric in momentum $\bk$ and band $n$, and antisymmetric in spin $s$, and therefore represented by the following Young diagram,
\begin{equation}
    \Yvcentermath1  \underbrace{ {\yng(2,2) \cdots\cdots \yng(2,2)} }_{N_s=2L^2}.
\end{equation}
The degeneracy of the irrep. $[N_s^2]$  can be calculated by Hook's formula \cite{Hofmann2022}:
\begin{equation}
  d_{[N_s^2]} = \frac{1}{12} (N_s+3)(N_s+2)^2(N_s+1) \propto N_M^4\,,
\end{equation}
giving rise to an entropy of
\begin{equation}
    S = k_\mathrm{B} \ln{d_{[N_s^2]}}\,.
\end{equation}

In addition to the unstrained data (Fig.~3 of the main text), the KIVC correlation function at finite temperature and the entropy results for finite strain, $\varepsilon=4\cdot 10^{-4}$, are shown in \figref{fig:entropy_ws}.
The KIVC correlation function shown in panel (a) is strongly enhanced around $1.14\leq \theta \leq 1.23$ at low temperature, $T\lesssim 10\mathrm{K}$. In addition, there is a single twist angle, $\theta=1.06^{\circ}$, which also shows enhanced KIVC correlations.
The disordered regime with $\theta > 1.23^{\circ}$ exhibits gapless electron excitations close to $\bk=K_M$ (see Fig.~1(b) of the main text) and is referred to as a Dirac semimetal. The disordered regime with $\theta < 1.14^{\circ}$ is the anisotropic semimetal with gapless single-particle excitations around $\bk=\Gamma_M$. Note that the Dirac points meander in the BZ due to the strain breaking the $C_3$ rotation symmetry, and meet at the BZ center for $\theta\approx1.06^{\circ}$. When the Dirac points meet, the dispersion becomes quadratic and the low energy density of states is enhanced, which might explain the reoccurrence of KIVC order for $\theta=1.06^{\circ}$. But we would like to stress that the Dirac cones of the anisotropic SM are not pinned to zero energy, inducing small electron and hole pockets which also have a non-zero density of states at the Fermi level. Currently, we cannot reach system sizes large enough to faithfully resolve those Fermi surfaces and determine whether the reoccurrence of the KIVC at $\theta=1.06^\circ$ is genuine or a finite size effect.

The entropy results are shown in \figref{fig:entropy_ws}(b-d) for three representative twist angle. 
For the anisotropic and the Dirac semimetal at $\theta=1.11^{\circ}$ and $\theta=1.4^{\circ}$, respectively, the entropy per unit cell smoothly evolves from $S/k_\mathrm{B}=8 \ln{2}$ at high temperatures down to $S/k_\mathrm{B}=0$ at low temperature.
At $\theta=1.2^{\circ}$ where the ground state is a KIVC state, the entropy is similar to the previous SM at high temperatures but exhibits a narrow ``step'' around $T\approx 30 \mathrm{K}$ with an entropy per unit cell close to $k_B \ln{\binom{8}{4}}$, before dropping sharply around $T=10\mathrm{K}$ to $S/k_\mathrm{B}<L^2\ln(2)$.
The intermediate plateau is consistent with a Hubbard-like regime where charge fluctuations are quenched, and the sharp drop is due to the crossover into the quasi long-ranged KIVC ordered phase.

\section{Analytic spectral function in fluctuating moment regime}
\label{app:AnalyticSpectra}

In this section we review the analytic spectral function calculations quoted in the main text. We will follow the treatment in Refs. \cite{Ledwith2025,ledwith2025a} for concreteness; see also Ref. \cite{Hu2025} for a similar calculation in the Song-Bernevig model and Refs.\cite{vituri2026controlled,nosov2026controlledexpansioncorrelatedelectrons,hu2026twisted,wei2026lifetime} for the development of a systematic order-by-order expansion in the parameter $s^2 \ll 1$.

\subsection{Small $s^2$ in TBG} 

We now review how to identify a small parameter $s^2\ll 1$ in TBG\cite{Ledwith2025}. We begin by restating the band projected model \eqref{eq:ham_QMC}
\begin{equation}
\begin{aligned}
    \hat{H} & = \sum_{\bk} c^\dag_{\bk} h^{\rm sp}_\bk c_\bk + \hat{V}, \\
    \hat{V} & = \frac{1}{2} \sum_\bq \overline{\rho}_\bq V_\bq \overline{\rho}_{-\bq}
    \end{aligned} 
    \label{eq:analytic_ham_start}
\end{equation}
where $h^{\rm sp}_\bk$ is the (renormalized) single particle dispersion matrix. Because the TBG bands are topological, they do not admit a symmetric localized Wannier basis. Instead, it is often convenient to work in ``sublattice basis" or ``Chern basis," in which the two flat bands per spin per valley can be decomposed into a $C=+1$ and $C=-1$ band \cite{Bultinck2020}. These Chern bands have most of their weight on the $A$ and $B$ sublattices respectively in valley $K$, and vice versa in valley $K'$. The single particle dispersion acts as a tunneling between the two Chern sectors. In this Chern basis, the form factors $\Lambda_{\bq}(\bk)$ that enter the projected densities $\overline{\rho}_\bq$ have an approximate $\U(4) \times \U(4)$ symmetry, originating from the strong sublattice polarization. Here each $\U(4)$ corresponds to spin-valley rotations in each $C=\pm$ sector. In the strong coupling limit, a large space of generalized quantum Hall ferromagnets obtained by occupying bands in the Chern basis are exact ground states at charge neutrality. Symmetry breaking terms and dispersion can be added perturbatively to split the ground state subspace, and they select the KIVC state at charge neutrality\cite{Bultinck2020,lian2020tbg,Vafek2020}. However, these arguments apply only to the ground state at integer filling and do not generalize to thermal states for example.

To make analytical progress for $T>0$, and understand the extensive entropy observed experimentally\cite{Rozen2021,Saito2021,zhangHeavyFermionsMass2025}, Ref. \cite{Ledwith2025} introduced a small parameter $s^2 \ll 1$. This was done by deriving a simple ansatz for the TBG Bloch wavefunctions based on the limit of concentrated charge density at the AA sites. The resulting wavefunctions have parametrically concentrated Berry curvature and matched the TBG wavefunctions to within $\sim 5\%$ error for a fixed value of the small parameter $s \approx 0.25$~\cite{Ledwith2025}. 
In the $\gamma = +1$ Chern sector they are given by 
\begin{equation}
  u_{\bk,\gamma=+}(\br) = \lambda_{\bk,\gamma=+} \sum_{\boldsymbol{R}}\left(1+\frac{(z-R)/\delta}{i k/s}\right) w_{\rm AA}(\br-\boldsymbol{R}).
    \label{eq:analyticalWfs}
\end{equation}
Here $z=x+iy, k=k_x+ik_y,$ and $R = R_x + i R_y$. The vectors $\bR$ are the moir\'e lattice vectors and the Bloch momentum $\bk$ is assumed to be in the first BZ. The function $w_{AA}(\br) = \frac{e^{-\frac{\abs{\br}^2}{4\delta^2}}}{\sqrt{2\pi \delta^2}} \chi_0$ is a Gaussian peak around the $\br = 0$ AA site, alongside a constant layer-space spinor $\chi_0$ that does not affect the form factors. The wavefunctions in a fixed valley and Chern sector are sublattice polarized according to the relation $\gamma_z = \sigma_z \tau_z$\cite{Bultinck2020}. The wavefunctions in the other Chern sector, $\gamma = -$ are related by $C_2 \mathcal{T}$, or $u_{\bk-} = \overline{u_{\bk+}(-\br)}$. We have assumed the $\U(4)$ symmetry of the wavefuntions within each Chern sector due to sublattice polarization, though we allow for $h^{\rm sp} \neq 0$. 

The modulus of the prefactor $\lambda_{\bk \gamma}$ normalizes the wavefunction while its phase defines the gauge in momentum space. Since the wavefunctions are those of a Chern band, we cannot choose a smooth and periodic gauge. We choose
\begin{equation}
  \lambda_{\bk +} = \lambda_{\bk -} = \lambda_\bk = \frac{\abs{k}}{\sqrt{\abs{k}^2 + 2s^2} }
  \label{eq:gaugechoice}
\end{equation}
unless stated otherwise. This puts the gauge singularity at the $\Gamma$ point. While we can work in any gauge, Eq. \eqref{eq:gaugechoice} makes the wavefunctions $u_{\bk}(\br)$ nearly $\bk$-independent for $\bk$ away from $\Gamma$. For such momenta, \eqref{eq:analyticalWfs} resemble the wavefunctions of a trivial band consisting of exponentially localized states at the AA sites. The Berry curvature associated with the band topology is concentrated in the region $|k|\sim s$. Our approximation uses the fact that this is a small area of phase space: $s^2 \ll 1$.

Wavefunctions like \eqref{eq:analyticalWfs} can emerge from several types of microscopic models. Indeed a similar wavefunction appears in the topological flat bands of the Song Bernevig model as a mixed-valence hybridization of $``f"$ and $``c"$ electrons (see SI of Ref. \cite{Song2022}).  We note that the $\bk$-dependence in \eqref{eq:analyticalWfs} has discontinuities at the unit cell boundaries that are small in $s$. 
The discontinuities originate from expanding a periodic function around each BZ center\cite{Ledwith2025}. This function depends on the details of the microscopic model with concentrated charge density (e.g. the BM model). 
The Song Bernevig model has a similar discontunuity that depends on the UV cutoff for $c$-electrons.
For convenience, in plotting figures we replace $1/k$ in \eqref{eq:analyticalWfs} with the Weierstrass zeta function $\zeta(\bk)$\cite{haldane2018modular}. This choice emerges naturally if \eqref{eq:analyticalWfs} are realized from a Dirac particle in a periodic and inhomogeneous magnetic field. All our conclusions will be unaffected by this choice; it only enters observables at order $s^2 \ll 1$ and is subleading to other corrections that are $\propto s^2 \log s^{-1}$ \cite{Ledwith2025}.

The Bloch wavefunctions \eqref{eq:analyticalWfs} define the interacting Hamiltonian \eqref{eq:analytic_ham_start} through the form factors $[\Lambda_\bq(\bk)]_{\alpha \beta} = \braket{u_{\alpha,\bk+\bq}}{u_{\beta,\bk}}$. These can be calculated straightforwardly by computing the requisite Gaussian integrals as in Appendix A of Ref. \cite{Ledwith2025}. Due to the small phase space $s^2 \ll 1$, we will only need the simplified expression
\begin{equation}
  \begin{aligned}
  \Lambda^{\gamma \gamma'}_{\bk,\bk+\bq\to*} & = \lambda_\bk \xi_\bq \begin{pmatrix}1 + \frac{\ov{q} \delta}{\ov{k}/s} & 0 \\ 0 & 1 + \frac{q \delta}{k/s}   \end{pmatrix}_{\gamma \gamma'}, \\ 
  \xi_\bq  & = e^{-\frac{|q|^2 \delta^2}{2}}.
  \end{aligned}
    \label{eq:simplifiedformfactor}
\end{equation}
Here we assume that $k$ is in the first BZ. We assume $\bk + \bq$ is far from $\Gamma$ (in any BZ), which we indicate through the notation $\bk + \bq \to *$. We also used a compact notation $\Lambda_\bq(\bk-\bq) = \Lambda_{\bk,\bk+\bq}$ and will continue to do so in the next subsection.

The single particle disperion has the form\cite{Ledwith2025}
\begin{equation}
  \begin{aligned}
  [h^{\rm sp}_\bk]_{\gamma \gamma'}&  = E_{\rm BM}(1-\abs{\lambda_\bk}^2)\begin{pmatrix} 0 & e^{-2 i \theta_\bk} \\ e^{-2 i \theta_\bk} & 0  \end{pmatrix}_{\gamma \gamma'} \\
  & + E_{\rm str}\abs{\lambda_\bk}^2 [\gamma_x]_{\gamma \gamma'}.
\end{aligned}
  \label{eq:spdisp_supp}
\end{equation}
The first term comes from tuning $\theta$ away from the magic angle; it leads to the splitting $\pm E_{\rm BM}$ at the $\Gamma$ point. The second term comes from heterostrain (taken along the $x$ axis without loss of generality here) and leads to a splitting $\pm E_{\rm str}$ in most of the BZ except for the $\sim s^2$ area region around the $\Gamma$ point.

\subsection{Green's function and Schwinger-Dyson Equation}

We now set up notation and discuss a general identity that we will use to compute the electron spectral function. All results stated in this subsection are fully general and do not rely on $s^2 \ll 1$. 

We will need to compute the Green's function of more general fermionic operators $d_\bk$ which may be multibody. We use the notation
\begin{equation}
  G^{d_1 d_2}_{\bk,n} = (d_{1\bk},d^\dag_{2 \bk})_n = -\int_0^\beta d\tau e^{i \omega_n \tau}  \langle \mathcal{T} d_{1\bk\tau} d^\dag_{2 \bk} \rangle,
\end{equation}
where $d_{\bk \tau} = e^{\tau \hat{H}} d_\bk e^{- \tau \hat{H}}$, $\omega_n = 2\pi(n+\frac{1}{2})$ is a fermionic Matsubara frequency, and $\mathcal{T}$ denotes imaginary time ordering. The bracket expression is useful for manipulations that rely on the linearity of $G^{d_1 d_2}$ in the operators $d_1$ and $d_2^\dag$.

The electron Green's function,
\begin{equation}
  G_{\bk,n} = G^{cc}_{\bk,n} = \left[ i \omega_n - h_\bk^{\rm sp} - \Sigma_{\bk,n} \right]^{-1},
  \label{eq:generalelectronGreensfunction_supp}
\end{equation}
follows from the computation of the self energy $\Sigma_{\bk,n}$. We will use a general identity for the self-energy, which we describe in detail after stating it.It has the form
\begin{equation}
  \begin{aligned}
    \Sigma_{\bk,n} = h^{\rm mf}_{\bk} + [G^{OO}_{\bk,n}]_{\rm 1PI}. \\
  \end{aligned}
  \label{eq:generalSigmaMFBeyond}
\end{equation}

The second term will be of primary importance to us. We will pause to briefly discuss the the first term, which is the frequency-independent ``Hartree-Fock'' contribution to the self energy:
\begin{equation}
  \begin{aligned}
    h^{\rm mf}_{\bk} & = h^{\rm H}_{\bk} + h^{\rm F}_\bk, \\
    h^{\rm H}_\bk & = \frac{1}{2A}\sum_\bG V_\bG \Lambda_{\bk,\bk+\bG} \sum_{\bk'} \tr \Lambda_{\bk',\bk'-\bG} Q_{\bk'}, \\ 
    h^{\rm F}_\bk & =  - \frac{1}{2A}\sum_\bq V_\bq \Lambda_{\bk,\bk+\bq} Q_{\bk + \bq} \Lambda_{\bk+\bq,\bk},
  \end{aligned}
  \label{eq:meanfieldcontribution}
\end{equation}
where $Q_\bk$ is defined through the Hartree-Fock order parameter $P_{\bk\alpha \beta} = \frac{1}{2}(1 + Q_{\bk\alpha \beta}) = \langle c^\dag_{\bk \beta} c_{\bk \alpha} \rangle $. We emphasize that we do not assume a mean-field state: for a mean field state Eq. \eqref{eq:meanfieldcontribution} describes the full self-energy.

The second term,
\begin{equation}
  \begin{aligned}
    [G^{OO}_{\bk,n}]_{\rm 1PI} & = G^{OO}_{\bk,n} - [G^{OO}_{\bk,n}]_{\rm 1PR},  \\
    & = G^{OO}_{\bk,n} - G^{Oc}_{\bk,n} [G^{cc}_{\bk,n}]^{-1} G^{cO}_{\bk,n}, \\
    O_\bk & = -[\hat{V},c_\bk] = \frac{1}{2A} \sum_\bq V_\bq \Lambda_{\bk,\bk+\bq}\{ c_{\bk + \bq},  \overline{\rho}_{\bq} \},
\end{aligned}
  \label{eq:beyondmeanfield1PI}
\end{equation}
corresponds to beyond mean field effects. The subscript 1PI (1PR) stands for one particle irreducible (reducible). The 1PR correlator corresponds to the parts of $G^{OO}$ mediated by a single-particle intermediate state, whereas the 1PI correlator subtracts off such contributions. The inverse electron Green's function in $[G^{OO}]_{\rm 1PR}$ corrects for the double counting of the intermediate $c$-fermion line: one from the ingoing line of $G^{Oc}$ and one from the outgoing line of $G^{cO}$ (see Ref. \cite{Ledwith2025} for a pedagogical derivation of this, both perturbatively and through the 1PI effective action). The contribution \eqref{eq:beyondmeanfield1PI} can therefore be understood as the contributions to the electron self-energy that are associated with scattering to strictly multi-particle states. These can then be naturally understood as the beyond mean-field contributions, as the mean-field contributions are precisely those that scatter electrons to other electrons through the mean-field potential.

We refer readers to Ref. \cite{ledwith2025a} for a proof of Eq. \ref{eq:generalSigmaMFBeyond}, and \cite{Ledwith2025} for the equivalent identity in a Hubbard-Stratonovich framework.
We emphasize again that all statements above are exact. We now use small $s^2$ to evaluate \eqref{eq:beyondmeanfield1PI} for the symmetric Mott state.

\subsection{Evaluation of Mott Self-Energy Using $s^2 \ll 1$}

To begin, we note that all diagrams contributing to the 1PI correlator \eqref{eq:beyondmeanfield1PI} remain connected after cutting any internal electron line. Thus, the electron momentum $\bk$ is always split between at least two internal lines, e.g. $\bk-\bq$ and $\bq$, with the relative momenta $\bq$ integrated over as a loop integral. For most values of $\bq$, these internal lines, and the internal vertices that they connect to, have momenta far from the $O(s^2)$ region near $\Gamma$. Thus, to leading order in $s^2$, the internal parts of diagrams contributing to the 1PI correlator \eqref{eq:beyondmeanfield1PI} can be taken to be away from $\Gamma$. For such momenta, $\hat{H}$ acts equivalently to the effective classical Hamiltonian
\begin{equation}
  \begin{aligned}
  \hat{H}_{\rm cl} & = \sum_\bR c^\dag_\bR h^{\rm sp}_* c_\bR +  \sum_{\bR,\bR'} U(\bR-\bR') \overline{n}_\bR \ov{n}_{\bR'}, \\
  U(\bR) & = \frac{1}{2A}\sum_\bq V_\bq \abs{\xi_\bq}^2 e^{-i \bq \cdot \bR}, \quad U(0) = U
\end{aligned}
  \label{eq:Hclassicalsupp}
\end{equation}
which is obtained from $\hat{H}$ through $h^{\rm sp}_\bk \to h^{\rm sp}_*$ and $\Lambda_{\bk,\bk+\bq} \to \Lambda_{\bk\to*,\bk+\bq\to*} = \xi_\bq$. Indeed, $\hat{H}$ and $\hat{H}_{\rm cl}$ have equivalent matrix elements when all momenta are taken to be far away from the small $\Gamma$ region. See Ref. \cite{Ledwith2025} for a pedagogical treatment of the above arguments, including the evaluation of simple example diagrams.

We are primarily interested with the $T \gg Us^2$ symmetric state with fluctuating moments, and we study the $E_{\rm str} = 0$ case for simplicity. This corresponds to the Mott semimetal discussed in the main text. Nonzero $E_{\rm str}$ causes a crossover to the anisotropic semimetal, while for $T\to 0$ one should obtain a KIVC state. We comment on how these other states can be understood in small $s^2$ at the end. We therefore have $h^{\rm sp}_* = 0$, such that the flavor-moments associated with \eqref{eq:Hclassicalsupp} are non-magnetized. Since we are also at charge neutrality, the Hartree-Fock order parameter associated to \eqref{eq:Hclassicalsupp}, which coincides with $Q_{\bk \to *} = Q_*$, vanishes. Then, to leading order in $s^2$, the Hartree and Fock terms \eqref{eq:meanfieldcontribution} vanish, as $Q \approx Q_*$ in most of the BZ. 

We therefore need to evaluate
\begin{equation}
\begin{aligned}
    [G^{OO}_{\bk,n}]_{\rm 1PI} & = \frac{1}{2A}\sum_{\bq} V_\bq \Lambda_{\bk,\bk+\bq} \xi_\bq \, \frac{1}{2A}\sum_{\bq'} V_{\bq'} \Lambda_{\bk+\bq',\bk} \xi_{\bq'} \\
    & \times\left[ \left( \{ \ov{n}_{\bq}, c_{\bk + \bq} \}, \{ \ov{n}_{-\bq'}, c^\dag_{\bk + \bq'} \} \right)_n  \right]_{\rm 1PI},
    \label{eq:factoroutlamks_supp}
    \end{aligned}
\end{equation}
with $\overline{n}_\bq = \sum_{\bk',\alpha} \left(c^\dag_{\bk'+\bq,\alpha} c_{\bk',\alpha} - \frac{1}{2}\delta_{[\bq],0}\right)$, where $[\bq]$ denotes the part of $\bq$ in the first BZ. We replaced the form factor $\Lambda_{\bk'+\bq,\bk'}$ inside the $\overline{\rho}_\bq$ part of $O_\bk$ with $\xi_\bq$ since both $\bk'$ and $\bq$ are summed over, and similarly for $O^\dag_\bk$. The $\Lambda_{\bk,\bk+\bq}$ and $\Lambda_{\bk+\bq',\bk}$ factors correspond to external vertices in self-energy diagrams and are associated with the electron leaving and re-entering the small $\Gamma$-point region respectively. As discussed above, to leading order in $s^2$ we can assume that no other lines or vertices depend on the small $\Gamma$ region. 

It follows that we can evaluate the bracketed 1PI correlator using the classical Hamiltonian \eqref{eq:Hclassicalsupp}. For $T \ll U$, the density matrix of \eqref{eq:Hclassicalsupp} has frozen charge fluctuations (it is annihilated by $\ov{n}_\bR = n_\bR - 4$) and random flavor occupations at each site. Its Green's function is purely local, and therefore momentum independent, and takes the form
\begin{equation}
  \begin{aligned}
  G^{cc}_{*,n} & = G^{cc}_{\bR,n} = G^{+}_n + G^{-}_n \\
  G^{\pm}_n & = \frac{\frac{1}{2}}{i\omega_n \mp U},
  \end{aligned}
  \label{eq:classicalgreensfn}
\end{equation}
where $G^+$ ($G^-$) corresponds to electron-like (hole-like) excitations associated with creating a site with $\ov{n}_\bR = +1$ $(-1)$, thereby costing an energy $U(0)=U$ in \eqref{eq:Hclassicalsupp}. The factor of $\frac{1}{2}$ is the probability for the excitation to not be Pauli blocked. We chose zero chemical potential here since we work exactly at charge neutrality in the QMC.

The evaluation of the correlator in Eq. \eqref{eq:factoroutlamks_supp} can be similarly obtained from Fourier transforming to real space and noting that the $\ov{n}_\bR$ factors sit between the fermion operators, otherwise they annihilate the density matrix, and give a positive (negative) sign for electron-like (hole-like) excitations respectively:
\begin{equation}
  \begin{aligned}
  \Xi_n & = \left[ \left( \{ \ov{n}_{\bq}, c_{\bk + \bq} \}, \{ \ov{n}_{-\bq'}, c^\dag_{\bk + \bq'} \} \right)_n  \right]_{\rm 1PI}, \\
 & = G^+_n + G^-_n + (G^+_n - G^-_n)(G^+_n + G^-_n)^{-1}(G^+_n - G^-_n) \\
 & = \frac{1}{i\omega_n}.
\end{aligned}
  \label{eq:evaluated1PIcorrelator}
\end{equation}

The sum over $\bq$ and $\bq'$ in \eqref{eq:factoroutlamks_supp} can now be evaluated. We have
\begin{equation}
    \frac{1}{2A}\sum_{\bq} V_\bq \Lambda_{\bk,\bk+\bq} \xi_\bq = \frac{1}{2A}\sum_{\bq} V_\bq  \lambda_\bk \abs{\xi_\bq}^2 =  \lambda_\bk U,
\end{equation}
where we used that the linear in $\bq$ part of the form factor vanishes in the sum due to $V_\bq =V_{-\bq}$ and $\xi_\bq = \xi_{-\bq}$. The term with the sum over $\bq'$ is the same. We therefore arrive at the Mott semimetal self energy,
\begin{equation}
  \Sigma_{\bk,n} = [G^{OO}_{\bk,n}]_{\rm 1PI} = U^2 \abs{\lambda_\bk}^2 \Xi_n = \frac{U^2 \abs{\lambda_\bk}^2}{i \omega_n}.
\end{equation}
and the resulting electron Green's function
\begin{equation}
  \begin{aligned}
    G^{cc}_{\bk,n} & = \frac{1}{i \omega_n - h^{\rm sp}_\bk - \abs{\lambda_\bk}^2 \frac{U^2}{i\omega_n}} \\
    & = \frac{P_+}{i\omega_n - E_{\rm BM}(1-\abs{\lambda_\bk}^2) - \abs{\lambda_\bk}^2 \frac{U^2}{i\omega_n}} \\
    & + \frac{P_-}{i\omega_n + E_{\rm BM}(1-\abs{\lambda_\bk}^2) - \abs{\lambda_\bk}^2 \frac{U^2}{i\omega_n}},
  \end{aligned}
  \label{eq:electronGreensfnBM}
\end{equation}
where $P_+$ and $P_-$ project onto the positive and negative energy bands of $h^{\rm sp}_\bk$ respectively (with $E_{\rm str} = 0$). For each BM band there are two quasiparticle branches, associated with the quadratic equations $\omega(\omega \mp E_{\rm BM}(1-\abs{\lambda_\bk}^2)) = \abs{\lambda_\bk}^2 U^2$. While the branches converge as $\bk \to *$, for $\bk \to 0$ one has $\omega \to E_{\rm BM}$ while the other has $\omega \to 0$. The branch with $\omega \to 0$ has zero quasiparticle residue at $\bk = 0$ because the self-energy pole causes the Green's function to vanish at zero frequency. These features can be seen in the Mott Semimetal spectral lines plotted in the main text and are consistent with the QMC spectra that they are plotted with (for twist angles not too far from the magic angle).

The vanishing quasiparticle residue is suggestive of an excitation orthogonal to the electron that mixes with the electron for nonzero $\bk$. In Ref. \cite{ledwith2025a} (see also Refs. \cite{zhao2025ancillatheorytwistedbilayer,zhao2025mixed} for treatments using an ancilla model) it was shown that this excitation can be created through a relatively simple ``trion'' operator, $F_{\bk \alpha} = \sum_\bR \{c_{\bR \alpha} , \ov{n}_\bR \} e^{-i \bk \cdot \bR}$. The Green's function in the full electron-trion space is
\begin{equation}
  \mathcal{G}^{-1}_{\bk,n} = \omega - \begin{pmatrix} h^{\rm sp}_\bk & U \lambda_\bk \\ U \lambda_\bk & 0 \end{pmatrix},
  \label{eq:electrontrion}
\end{equation}
where the off-diagonal terms correspond to electron-trion tunneling. Projecting \eqref{eq:electrontrion} onto the electron degree of freedom $\begin{pmatrix} 1 & 0 \end{pmatrix}^T$ consistently recovers the electron Green's function \eqref{eq:electronGreensfnBM}.

\subsection{Anisotropic semimetal and KIVC Insulator}

We close by commenting on the anisotropic semimetal and KIVC states. For $E_{\rm str}$ nonzero, one obtains a crossover\cite{ledwith2025a} between the Mott semimetal ($E_{\rm str} \ll T$) and the anisotropic semimetal ($E_{\rm str} \gg T$). The crossover is driven by the magnetization of the $\gamma_x$ moments in \eqref{eq:Hclassicalsupp}.
The anisotropic semimetal obtained from $E_{\rm str} \gg T$ and $s^2 \ll 1$ is a Slater determinant with $Q_* = -\gamma_x$ and dispersion $\Sigma_{\bk,n} = h^{\rm F}_\bk = U\abs{\lambda_\bk}^2 \gamma_x$ (the 1PI correlator vanishes for $Q_*^2 = 1$). This leads to a quadratic band touching at $\Gamma$ for $E_{\rm BM} = 0$ and two Dirac cones near $\Gamma$ for $E_{\rm BM} \neq 0$. 

The KIVC state at $T \ll Us^2$ requires going to higher order in $s^2$ and accounting for exchange interactions $\propto Us^2$\cite{Ledwith2025}. Its emergence can be understood at a high level by adding a mean-field exchange proportional to the KIVC order parameter\cite{Bultinck2020} by hand: $h_{\rm KIVC} \sim Us^2 \eta_x \gamma_z$, where $\eta_x$ flips valley but not Chern sector. For $T \ll Us^2$ this is sufficient to fully magnetize the flavor moments of \eqref{eq:Hclassicalsupp} and give $Q_* = \eta_x \gamma_z$. This $Q_*$ in turn generates a \emph{gapped} dispersion through $\Sigma_{\bk,n} = h^{\rm F}_\bk = U \abs{\lambda_\bk}^2 + U_\Gamma(1-\abs{\lambda_\bk}^2)$, where
\begin{equation}
  U_\Gamma = \frac{1}{2A} \sum_\bq V_\bq \abs{\xi_\bq}^2 (\tfrac{1}{2}\abs{q}^2 \delta^2) 
  \label{eq:Ugamma}
\end{equation}
is the energy scale associated with exchange interactions between the $\Gamma$ point and the rest of the BZ (or the $c$-electron orbital and the $f$-electron orbital in the Song-Bernevig basis\cite{Song2022}). For $V_\bq \propto |q|^{-1}$ we have $U_\Gamma = \frac{U}{4}$. We saw above that these exchange interactions, generated from the $q$-linear parts of the form factor \eqref{eq:simplifiedformfactor}, destructively interfere for the inter-Chern nematic semimetal state and the flavor-incoherenent Mott semimetal, such that they do not generate a gap at $\Gamma$ for these states.

\end{document}